\documentclass[12pt,preprint]{aastex}

\shorttitle{Active region transition region loop populations}
\shortauthors{Ugarte-Urra}

\begin{document}


\title{Active region transition region loop populations\\
and their relationship to the corona}

\author{Ignacio Ugarte-Urra\altaffilmark{1}, Harry P. Warren and David H. Brooks\altaffilmark{1,2}}
\affil{Space Science Division, Code 7670, Naval
       Research Laboratory, Washington, DC 20375}
\altaffiltext{1}{College of Science, George Mason University; 4400 University Drive; Fairfax, VA 22030}
\altaffiltext{2}{Present address: Hinode Team, ISAS/JAXA, 3-1-1 Yoshinodai, Sagamihara, Kanagawa 229-8510, Japan}


\begin{abstract}
The relationships among coronal loop structures at different temperatures is not settled.
Previous studies have suggested that coronal loops in the core of an active region are 
not seen cooling through lower temperatures and therefore are steadily heated. If loops were 
cooling, the transition region would be an ideal temperature regime to look for a signature of
their evolution. The Extreme-ultraviolet Imaging Spectrometer (EIS) 
on {\it Hinode} provides monochromatic images of the solar transition region and corona at an 
unprecedented cadence and spatial resolution, making it an ideal instrument to shed light
on this issue.  Analysis of observations of active region 10978 taken in 2007 December 
8 -- 19  indicates that there are two dominant loop populations in the active region: core 
multi-temperature loops that undergo a continuous process of heating and cooling in the full 
observed temperature range $0.4-2.5$ MK and even higher as shown by the X-Ray Telescope (XRT); 
and peripheral loops which evolve mostly in the 
temperature range $0.4-1.3$ MK. Loops at transition region temperatures can reach heights of
150 Mm in the corona above the limb and develop downflows with velocities in the range of 
$39-105 \, \rm km\,s^{-1}$.
\end{abstract}


\section{Introduction}
Loop structures in the Solar corona have been observed at temperatures
spanning several thousand to several million K. Lately considerable effort has gone
into understanding the relationship among these structures. Despite these 
efforts, we have not yet developed a fully coherent 
picture of how structures at different temperatures are related. One of the goals is to 
understand whether these are independent structures governed by different 
heating mechanisms with different temporal and spatial characteristic scales, 
or whether they are just snapshots of the same structures in different stages of 
their evolution. 

We do know that active regions do not look exactly the same in different 
temperature regimes. The core of the active region is generally dominated 
by hot, high density loops. Electron temperatures in an active region core are
typically near 3 MK \citep[e.g.][]{saba1991,brosius1997}. The footpoints of 
these hot loops form the ``moss'', which is the bright network pattern observed 
in emission lines formed near 1 MK \citep{berger1999}.
The periphery of the active region is generally dominated by longer loops with 
lower temperatures \citep{delzanna2003a}. 

The relationship between the hot core loops and the ``warm'' loops has not been completely 
established. In fact, it has been proposed that they are largely independent.
\citet{antiochos2003} argued that soft X-ray loops in the  cores of active 
regions are not seen cooling to lower temperatures in filter images, therefore implying a steady 
source of heating \citep[see also][]{martens2008}, and some success has been 
achieved in modeling such a scenario \citep{winebarger2008, warren2008}. It is evident from other
observations, however, that  many soft X-ray active region loop structures cool down and 
are observed in the EUV at 1 -- 2.5 MK \citep{winebarger2005,ugarte-urra2006,warren2007}. The open 
question  is which is  the dominating component and what role do observational 
constraints, such as spatial resolution, play in that apparent steadiness. No 
observational study has given a fully satisfactory answer to this question yet.
 
If coronal loops are cooling, it seems natural to look for signatures of the 
evolution at lower temperatures.
The relationship of the hot and warm loops to structures at temperatures between 
0.1 and 1 MK, traditionally called transition region temperatures, is however
unclear. To the complex nature of the transition region \citep[e.g.][]{mariska1992} 
we have to add the fact that even though images of the corona have been routinely 
taken by Soft X-ray (e.g. SXT/{\it Yohkoh}, XRT/{\it Hinode}) and Extreme Ultra-Violet 
(e.g. EIT/{\it SOHO}, {\it TRACE}, 
EUVI/{\it STEREO}) imagers for many years now, that has not been the case for temperatures 
between 0.1 and 0.8 MK. Our understanding of this important and very dynamic 
component of the atmosphere is mostly based on spectroscopic observations, which 
often lack simultaneous good spatial and temporal sampling. Spectroscopic 
images take several minutes, sometimes hours, to build up, while sit-and-stare 
time series lack context. The latest best efforts for their high spectral and 
spatial resolution and sensitivity have been made with the CDS and SUMER spectrometers 
on board {\it SOHO}. Observations have shown that loop structures are common at 
transition region temperatures and that they reach heights (100 Mm) usually associated 
with coronal plasmas \citep[e.g.][]{brekke1999}. These loops are nearly complete 
or consist of long segments, suggesting isothermality \citep{fredvik2002, chae2000}. 
Sometimes they form condensations, also seen by {\it TRACE} \citep{schrijver2001}, and 
they are characterized by their short-term variability on the order of tens of minutes
\citep{kjeldseth-moe1998, fredvik2002}.

These studies, however, failed to reach a definite conclusion on the relationship of 
these structures to higher temperature coronal plasmas 
\citep{brekke1999}. On the one hand, the co-spatiality and co-temporality 
of loops at multiple temperatures was reported \citep[e.g.][]{kjeldseth-moe1998,spadaro2000,schmelz2006}.
On the other, authors like \citet{fludra1997} have stated that the evolution of plasma with 
temperatures under 0.6 MK is independent of the evolution of the coronal plasma. In 
similar terms, \citet{matthews1997} and \citet{harra1999} have also concluded that high lying 
cool loops are not the result  of the cooling of hot loops, but rather that they are
separate co-existing entities. The lack of connection between active region 
emission in the transition region and the corona was also discussed by other authors
\citep{feldman1994,landi2004}. Despite their conclusions, some of these studies 
acknowledged the difficulty of interpreting simultaneous images at different 
temperatures, which leaves open the possibility that these transition region loops 
may be the result of the cooling of hot structures \citep[e.g.][]{strong1996,harra2004}. 

At this crossroads, we present an initial analysis of transition region imaging from
the Extreme-ultraviolet Imaging Spectrometer (EIS) on board {\it Hinode}, to clarify
some of the open questions raised by these earlier studies. In addition to the 
1\arcsec\ and 2\arcsec\ narrow slits, EIS also has a 40\arcsec\ slot. This slot 
allows EIS to image the Sun over relatively narrow ($\approx$1 \AA) wavelength ranges.
The broad range of temperatures in the EIS wavelength ranges provide monochromatic 
images in lines ranging from \ion{Mg}{6} to \ion{Fe}{16}. The EIS slot images are 
similar to those taken with the {\it Skylab} S082A instrument \citep{tousey1977}, except 
that the blending is significantly reduced: images subtend
1 \AA\ on the detector versus the 25 \AA\ in {\it Skylab}. The field of view is limited,
but larger fields of view can be imaged by stepping the slot across the region of interest.

Our main conclusions, in Section~\ref{sect:conclusions}, are that active regions 
have, at least, two main loop populations: the core loops that are energized to several 
million K and cool down to transition region temperatures; and the peripheral cool loops 
that have peak emissions in the range $0.4-1.3$ MK. The nature as well as the temporal and 
spatial characteristics of the heating of these structures still remain to be determined. 
More details on the loop results are given in Section~\ref{sect:looppopulation}. 
First, we introduce in Section~\ref{sect:observations} the EIS spectrometer, the set
of observations and why they are of interest.

\section{Observations}
\label{sect:observations}
\subsection{EIS observing and mode selection}
The EUV Imaging Spectrometer (EIS) \citep{culhane2007} on {\it Hinode} \citep{kosugi2007}
is a flexible spectrometer which can be used in three main modes of 
operation. Firstly, using the 1\arcsec\ or the 2\arcsec\ slits, it can 
simply retrieve a sequence of spectra of the same solar location to allow 
the study of its full spectral properties at consecutive times. 
Secondly, using the same slits and a scan motion it can retrieve a sequence 
of spectra at adjacent solar positions to study the full spectral properties 
of an extended region, up to $590\arcsec\times512\arcsec$. In the first mode, 
temporal resolution is gained at the expense of spatial extension and the 
opposite occurs in the second mode. Intermediate configurations 
allow a compromise between the two. In these two modes, the spectroscopic 
capabilities are normally prioritized over spatial and temporal sampling.

EIS also allows us to trade part of the spectral capabilities to improve 
the cadence and field-of-view coverage, giving the former three properties 
almost equal importance. This can be accomplished by using the 40\arcsec\ 
(also 266\arcsec) aperture, also called the slot. Each spectral line then produces 
its own 40\arcsec\ image. As a result the detector hosts as many images as 
there are spectral lines  in the wavelength range. Images produced by 
spectral lines with centroids closer than 40 spectral pixels (one spectral 
pixel corresponds to 1\arcsec\ on the Sun) will overlap and be difficult
to interpret. There are, however, many spectral lines that are
sufficiently isolated from neighboring strong lines. 
For those lines EIS can achieve  imaging capabilities 
 comparable to those of EUV imagers, at a greater spectral, 
and therefore temperature discrimination. Single exposures cover as much as
$40\arcsec\times512\arcsec$ and require shorter exposure times than 
the narrow slits, which allows better temporal sampling. The trade off to obtain 
these fast spectrally pure images is an overlap of spectral and spatial 
information and, therefore, the loss of a fully resolved spectrum that could be
subject to the usual spectral analysis: line fitting, etc.  Similar modes of 
operation have already been used in the past, e.g. {\it Skylab} spectroheliograms
\citep[e.g.][]{tousey1973} or slot images with the Coronal Diagnostic 
Spectrometer on board {\it SOHO} \citep[e.g.][]{ugarte-urra2004}. EIS's
novelty is its improved sensitivity, spectral coverage and spatial resolution.

\subsection{Active region NOAA 10978}
Interested in the relationships among loops at different temperatures and
aware of the unique capabilities of the EIS for spectral imaging, our aim 
was to take a fresh look at the evolution of active region loop structures in unprecedented
detail. Our goal was then to look at them in a similar way to that which has been done
with EUV imagers, but taking advantage of the temperature discrimination that 
the slot images provide. Fig.\ref{fig:bigpanel} shows the temperature 
response of the EIS slot images (color solid lines) in comparison to three 
{\it TRACE} and XRT passbands (dotted lines). 
The lines on the graph represent the temperature range where lies 95\% of the 
contribution funcion of the spectral lines\footnote{Using the CHIANTI atomic database 
\citep{landi2006}, version 5.2.} in the case of EIS and 95\% of the 
temperature response in the case of {\it TRACE} and XRT.

We therefore planned a 12 day run of observations 
of active region (AR) NOAA 10978 with long sequences of slot imaging, while 
the AR rotated from the East to the West limb. The study basic unit consisted 
of four $40\arcsec\times400\arcsec$ exposures of 15 seconds at four adjacent 
solar positions with a 5\arcsec\ overlap, resulting in a $140\arcsec\times400\arcsec$ 
region sampled every 70 seconds in eighteen different spectral lines. Spectral
lines were chosen for their spectral purity and temperature coverage, see 
Table~\ref{tab:obsinfo}, assessed in previous full CCD observations. It is worth 
noting the presence of three purely transition region lines: \ion{Mg}{6}
($\rm T \approx0.4 \,MK$), \ion{Mg}{7} and \ion{Si}{7} ($\rm T \approx0.6 \,MK$).
Studying the evolution of loop structures in 
spectrally pure images (isothermal plasma) of the transition region and the 
corona at a 70 s cadence should be sufficient to establish the, heretofore unclear,
relationship between plasmas at those temperatures and timescales.
By isothermal we mean a temperature response as narrow as the width of 
the line transition's contribution function \citep[e.g.][]{mariska1992}.

AR 10978 (Fig.~\ref{fig:bigpanel}) appeared over the East limb on December 4, 2007. It was a moderately 
sized active region that, on its passage towards the West limb, experienced 
moderate activity in terms of GOES fluxes. As shown in Fig.~\ref{fig:goes}, 
ten C-flares are associated with its development during the twelve 
days observing time. The figure also indicates (Greek symbols) the magnetic 
classification of AR 10978 as given by NOAA. The index gives a notion of 
activity in the active region \citep[e.g.][]{ireland2008}. 
\citet{dalla2007} investigated a sample of 2880 sunspot regions from the NOAA
Solar Region Summary catalog and found that 73\% of active regions have 
$\beta$ as maximum magnetic classification and 11\% reach $\beta\gamma$. 
As shown in Fig.~\ref{fig:goes} AR 10978 belongs to the latter group, i.e.
a bipolar sunspot group with more than one clear north-south polarity
inversion line.

The data were reduced using standard EIS software. The slot images,
similarly to narrow slit observations, experience a drift on the detector due
to the satellite orbital changes \citep{brown2007}. This is corrected via 
cross-correlation of consecutive images, which also removes the spacecraft jitter.


\section{Active region transition region emission and loop population}
\label{sect:looppopulation}
Our study of several hours a day for a 12 day observing sequence of multi-wavelength 
slot movies reveals that active regions can exhibit transition region contributions 
from two distinct loop populations: compact multi-temperature core loops and
cool and extended peripheral loops. Other significant contributors are active region
transition region brightenings and an unresolved low lying component. Although the 
individual contribution of all these parts has been reported before, 
\citep[see introduction papers and references therein, and][for EIS initial results]{young2007}, we will give
a further insight into their relationship with other structures seen at different
temperatures. Our emphasis will be towards the two distinct loop populations.

\subsection{Compact multi-temperature loops}
\label{sect:multitemp}
To the trained eye in EUV coronal loop observations, the most novel and noticeable 
features in the slot movies are transition region loops
characterized for being short lived and clearly defined along their full length 
(or a large portion of it). Loops with similar characteristics were already observed 
with CDS \citep{brekke1999}. They connect footpoints at both sides of the neutral line 
in the same way the hot coronal emission does. An easy proxy of their locii in the
current dataset is the envelope of the AR core and diffuse emission in the C\_poly 
XRT image in Fig.\ref{fig:bigpanel}. Therefore, one first thing to notice is that they
occur where hot emission ($\ge$2.5MK, i.e. \ion{Fe}{16}) is present.

Their relationship with structures at higher temperatures becomes evident when 
plotting lightcurves of the intensity of different spectral lines with different 
formation temperatures. The lightcurves indicate that these transition region loops
are the result of cooling of multi million degree plasma.
Fig.\ref{fig:loops_lightc} shows examples for December 11 (top half) and December 9 
(bottom half). The lightcurves, in normalized intensities, are shown for three 
different locations on the active region (boxes 1, 2 and 3). Different spectral lines
are indicated in different colors (see color coding on the figure). On top of 
the lightcurve panels we show context composite slot images for all these spectral 
lines at the time of maximum intensity in box 1, dashed lines on box 1 lightcurve panel. 
These slot images are available as movies in the online version of the paper.

The loop under box 1 on December 11 is a good example. 
The loop is first observed at temperatures $\ge 2.5$ MK (\ion{Fe}{16}) and slowly
cools down through all the spectral lines sampled by EIS. It takes approximately 65 
minutes to reach a maximum in intensity at transition region temperature (\ion{Si}{7})
and its identification is unambiguous at all temperatures. The other boxes show 
similar general patterns except for transient brightenings (e.g. box 3 on December 9 
13:30 UT) which will be discussed later on. 

Loops clearly seen at transition region temperatures can sometimes be difficult to 
identify in \ion{Fe}{15} -\ion{}{16} images due to the higher background and foreground
emission, interpreted as a larger population of loops at those temperatures at any
instant in time. This is better quantified by stating that the contrast of the loops
to the background emission ranges 1.4 -- 2.0 in \ion{Fe}{15}, 1.7 -- 4.0 in \ion{Fe}{12}
and 2.2 -- 5.8 in \ion{Si}{7}.

As seen in the two examples in Fig.\ref{fig:loops_lightc}, these transition region 
loops emit along their full length or a big fraction of it, suggesting isothermality
along the structures at any given instant. The lightcurves, however, show an evident overlap 
at different temperatures, which also suggests that the thermal distribution  has a
certain width. We present an emission measure analysis of this active region elsewhere 
\citep{warren2008b}. Results indicate that the thermal distribution along the 
line-of-sight has a typical width
of $\approx3\times10^5$ K. It is important to note here that we do not see transition 
region loops, as described in this section, that do not reach coronal temperatures.

{Lifetimes at transition region temperatures are on the order of tens of minutes. 
Table~\ref{tab:looplifetimes} gives lifetime estimates for a sample of several loops 
that were background subtracted in coronal and transition region lines. The lifetime
is defined as the full width at half maximum intensity. 
Evolution times are longer for coronal lines. This is also reflected in the smoothness 
of the lightcurves that increases for larger temperatures. The smooth envelopes of 
coronal emission can often be associated with several cooler events. This has been
seen before in different scenarios, and together with the fact that loops appear to live longer 
than their characteristic cooling time, has lead some authors to consider multiple 
threads as an important part of loop evolution at this resolution \citep[e.g.][]{warren2002,warren2003}.
 
The current dataset is one of the most complete in terms of temperature coverage for a 
loop evolution and this issue can be investigated with some detail. The left panel on 
Fig.~\ref{fig:tdecay} shows the estimated temperature decay for the loop that lies 
underneath box 1 on December 11. Each datapoint represents the time when the loop's 
lightcurve for that specific spectral line reaches half its peak intensity in the rising 
phase, and the temperature where we expect that spectral line to become observable. We 
arbitrarily define it as the largest temperature where the atomic contribution function 
of the spectral line is larger than half its maximum. The red dot-dashed 
line corresponds to the linear fit to those points. Even though an exponential fit can be
a fair approximation of the decay in the range 1.5 -- 3 MK, the linear fit reproduces
better the observations in the overall temperature range. 
The right panel in the figure shows in black the normalized
background subtracted intensities for that loop as function of time. The red dot-dashed
curves represent the variation in time of the spectral line contribution function assuming
the temperature changes indicated by the linear fit. The contribution function is proportional 
to the emissivity of the line, i.e. it is a proxy of the loop's lightcurve. This plot shows 
that the observed loop's lifetime can be very close to the expected one from the characteristic
cooling time given by the fits. This is the case for some of the cooler lines. There are large 
inconsistencies, however, for other lines that have slow intensity decay tails. It is not clear
therefore that lifetimes and lightcurves can be fully explained by the cooling rate. A study
of a statistically significant sample should give us more insight in whether the discrepancies
are systematic.


Another interesting note on these loops is that sometimes they recur, i.e. a loop 
with similar topology goes through several cycles of (heating and) cooling. An example 
is box 1 lightcurve on December 9. The behavior has been reported before \citep{shimizu1995,ugarte-urra2006}. 

These properties are true in the quietest period between days 8 and 12 
(see Fig.~\ref{fig:goes}), and become more evident when flaring activity steepens up 
on days 12, 13 and 14. During those days the active region is observed on-disk. 
Off-limb (December 19) the picture becomes more confusing, as the emission from loops 
in the core lies in the same line-of-sight as the legs of long peripheral loops described
in the next section. Furthermore, the contribution from foreground and background 
emission at 1-2 MK in the core is larger off-limb, as we integrate through twice the 
contribution: both legs versus just loop tops on-disk. Individual loops can still be 
pinpointed in the transition region lines, but their relationship to hotter temperatures 
is not as clear. This would explain partly why CDS studies, many off-limb, resulted in 
different conclusions. Another reason is that those studies did not have the
spectral coverage at the temporal cadence that EIS slot movies provide.

\subsection{Peripheral extended cool loops}
A second distinct population of loops is located at the periphery of the active region core.
In Fig.~\ref{fig:bigpanel} these loops are the ones that are visible in the lower left 
quarter of the {\it TRACE} 171 \AA\ image, but not at the same location in XRT. In EIS
slot images those loops are most prominent in \ion{Si}{7} and \ion{Mg}{7}, which
correspond to a temperature of 0.6 MK in ionization equilibrium, below the temperature 
of maximum response of the 171 \AA\ passband. Their cool nature was already diagnosed 
using the spectroscopic capabilities of CDS \citep{delzanna2003a,delzanna2003b} 
and more recently EIS \citep{young2007}. Those studies lack temporal information and justify
our investigation of the temporal evolution of the loops. Our analysis confirms that 
the main contribution to these loops over time is in the range 0.4 -- 1.3 MK. Unlike the 
loops described in Section~\ref{sect:multitemp}, these loops do not seem to be the result 
of cooling of $\ge$2.5 MK loops.
 
Fig.~\ref{fig:coolloops_lightc} shows lightcurves for representative locations on
two different days, December 9 (top) and December 10 (bottom). Notice there is a 
data gap of 2.4 hours on December 10 (filled pattern) due to EIS operational constraints.
The intensity lightcurves 
from loop structures at 0.4 and 0.6 MK (respectively in blue and brown colors, \ion{Mg}{6} 
and \ion{Si}{7}) are correlated.  Yet, in contrast to the lightcurves in Fig. ~\ref{fig:loops_lightc}, 
there is not such a clear cut correspondence between the intensity changes at transition region
temperatures and the changes in lines formed at higher coronal temperatures. 
The lightcurves suggest that loops at transition region and lower coronal temperatures 
evolve independently of the hotter structures.
In some instances, like December 9 box 2 between 13:00 UT and 14:00UT, the enhancements in 
the \ion{Si}{7} and \ion{Mg}{6} lightcurves (and movies) can be associated to earlier changes in 
\ion{Fe}{11} and \ion{Fe}{12}, suggesting that these loops could well reach temperatures up
to 1.3 MK. However, our analysis does not give any indication that these loops reach the 
formation temperatures of \ion{Fe}{14} -- \ion{Fe}{16}. Lightcurves of boxes 1, 2 an 3 on 
December 10 do show  in several instances a similar pattern of cooling as described in the previous section, however, 
a close inspection of the movies reveals that this is the result of line-of-sight contamination 
from multi-temperature loops and can not be interpreted as a characteristic evolution of the 
background peripheral loops.

The \ion{Si}{7} and \ion{Mg}{6} lightcurves reveal structuring within the smoother 
general trends, for example in box 3 (December 9). The time scales for these structures is on 
the order of tens of minutes and it is likely that some of these changes are 
oscillatory in nature. Further work is in progress to investigate it. 
The loops are clearly visible as fan structures
in {\it TRACE} 171 \AA\ and the presence of waves in this type of loops is a well documented 
phenomenon \citep[e.g.][and references therein]{demoortel2002}.


\subsection{Off-limb emission}
Emission at the coolest temperatures in the dataset is not just constrained to the footpoints, 
but it extends into coronal heights. That is best seen in  projection over the limb. 
Fig.~\ref{fig:loops_offlimb} shows an image of the active region on December 19 over the 
West limb. As pointed out earlier, off-limb images have a large line-of-sight contamination, 
mostly at coronal temperatures. In the cooler lines, some loops can be clearly isolated. Their 
legs extend as much as 150 Mm and more interestingly the intensity changes,
mostly in \ion{Mg}{6}, indicate the presence of downflows along the legs towards the 
surface. These downflows are also noticeable on-disk. On projection over the plane of the 
image, the velocities are in the range  $39-105 \, \rm km\,s^{-1}$. Fig.~\ref{fig:loops_flows}
shows various examples of downflows in three different loops. The plots represent the
change of intensity along the loop's axis in time, with the limb at the bottom. The intensities
have been smoothed in a boxcar of 2 pixels to reduce noise. Running differences are shown to 
enhance the downflow path. Asterisks indicate the maximum intensity in time at a particular position
along the axis. The velocities result from the linear fit to these points.
This phenomenon has been
referred to as coronal rain \citep{foukal1978} and it appears to be a rather common process  in active
regions interpreted in terms of catastrophic cooling of the loops \citep{schrijver2001,mueller2005,karpen2006}. 
Simultaneous \ion{Ca}{2} H images from the Solar Optical Telescope on board {\it Hinode}, show 
clumps of material continuously falling at chromospheric temperatures off-limb.

\subsection{Transition region brightenings and the unresolved component}
Sudden (few frames) brightenings are another contributor to transition region emission.
They tend to be point-like, but they can be elongated too. 
At the movies cadence (70 s) these events brighten up simultaneously 
at several temperatures. Sometimes they reach up to Fe XVI temperatures, some other 
times they seem constrained to the cooler spectral lines. Their impulsiveness does       
not allow to resolve their temperature evolution at this cadence, which makes them
distinct from the other discussed loop samples. One example is the sudden 
intensity increase seen in box 3 on December 9 13:26 UT at all temperatures. 
The magnetic changes associated with them appear different from the ones associated 
with the clearly defined loops described in previous sections \citep{brooks2008}. The 
dimensions of these events are often close to the spatial resolution of the spectrometer.
Earlier studies have already discussed the properties of active region transient events 
that reach coronal temperatures \citep[][and references therein]{shimizu1995, berghmans2001} and 
events that remain cooler 
\citep[namely explosive events, e.g.][]{perez1999}. We should stress however that 
the majority of transient events described by \citet[][]{shimizu1995} present a single
loop or multiple loop morphology and lifetimes which are also consistent with the 
loop population described in Section~\ref{sect:multitemp}. In fact, transient X-ray loops 
have been seen cooling in the  EUV \citep{ugarte-urra2006,warren2007}. 
Therefore, we believe that observations
with higher spatial and temporal resolution would be needed to investigate if these
sudden unresolved transient coronal brightenings are different in nature  \citep{brooks2008}
to the transient loops described in Section~\ref{sect:multitemp}.
  
Finally, and for completeness, there is also a transition region contribution from 
an unresolved component, which we will not discuss further in this paper. The origin
of this component could be associated with the footpoints of coronal loops, the thermal 
interface in a continuous stratified atmosphere \citep[e.g.][]{griffiths2000}, or to
unresolved cool structures disconnected from the corona \citep[e.g.][]{landi2004}.

\section{Discussion and conclusions}
\label{sect:conclusions}
Our investigation of monochromatic images and movies of the solar atmosphere 
taken with the Extreme-ultraviolet Imaging Spectrometer at an unprecedented 
cadence and spatial resolution gives new clues about the relationship 
between loop structures seen in different temperature regimes at different 
times. 

We find that loops in the core of the active region experience
a heat deposition that makes them reach temperatures at least as high as 2.5 MK, 
which is then followed by a cooling process down to 
transition region temperatures, as low as the temperature response of the coolest 
spectral line in the dataset (0.4 MK). The process can recur in a time-span of 
several hours. In fact, the temperatures reached by these loops are probably higher.
We have investigated the soft X-ray lightcurves (XRT) \citep[][]{golub2007} of a 
few examples and soft X-ray enhancements precede the EUV response. 
Fig.~\ref{fig:xrt_eis} shows the XRT C\_poly$/$Open and Open$/$Al\_thick filter combination
lightcurves for Box 1 on December 9 and 11. The peak in the temperature response 
for those filters is 8 and 12 MK respectvely.
This is in agreement with our
previous studies \citep{ugarte-urra2006, warren2007} and indicates that there
is a non-negligible component of loops in active regions that undergo a 
continuous heating and cooling process.

The loops emit along their full length, or a big fraction of it,
at all temperatures. Together with the fact that the thermal distribution has 
a typical width of $\approx3\times10^5$ K \citep{warren2008b} this suggests that
at any given instant in time, loops are isothermal structures, possibly
filamented, with a cross-field coherence. 

Previous studies \citep{fludra1997,matthews1997,harra1999} have suggested that 
there is a population of transition region loops reaching coronal heights 
evolving independently from the coronal plasmas. At the core we do not find 
evidence of loops at transition region temperatures that are not the result 
of cooling from multimillion degree plasma. The evolution of these cool loops is 
associated with the cooling of coronal structures. At the core's periphery, 
however, we do observe loops with their major contribution in the range 
$0.4-1.3$ MK and we do not find evidence in our datasets that these loops result 
from the cooling of multimillion degree plasmas. This result confirms previous 
spectroscopic diagnostics of the loops. 

These two populations are the dominant components of the coronal and transition
region emission during the full week passage of the active region and as such
have to be explained by any physical model. It remains to be proved, however, that 
these populations prevail in every active region. In fact, that it is probably not the
case, as the existence of a much steadier multi-million K loop population has 
been reported in the past \citep[e.g.][]{antiochos2003}. Given the proposed
scenarios for coronal heating, e.g. magnetic braiding and reconnection, it is 
likely that the level of magnetic complexity at the photospheric level is related to 
the soft X-ray and EUV response in the atmosphere.
We have shown that this active region has a magnetic classification ($\beta\gamma$) 
that would put it among a minority (11\%) of observed active regions \citep{dalla2007}.
Therefore, future investigations should address the relationship between loop
evolution and magnetic complexity in an attempt to identify and characterize the
various mechanism at play that result in the distinct loop populations and 
their evolution.

\acknowledgments
This work was supported by Hinode, a Japanese mission developed and launched by ISAS/JAXA, 
with NAOJ as domestic partner and NASA and STFC (UK) as international 
partners. It is operated by these agencies in co-operation with ESA 
and NSC (Norway).We are thankful to John Mariska and the anonymous referee for 
the careful reading of the manuscript and the comments that helped to improve it. 



\clearpage
\begin{deluxetable}{rlcc}
\tabletypesize{\footnotesize}
\tablecolumns{4}
\tablecaption{Spectral lines}
\tablewidth{0pt}
\tablehead{\multicolumn{2}{c} {Ion}  & Wavelength [\AA] & T [MK]}
\startdata
He     & \ion{}{2}  	& 256.3 &   0.05\\
Mg     & \ion{}{6}  	& 269.0 &   0.40\\
       & \ion{}{7}  	& 278.4 &   0.63\\
Si     & \ion{}{7}  	& 275.4 &   0.63\\
       & \ion{}{10} 	& 258.4 &   1.26\\
       & \ion{}{10} 	& 261.1 &   1.26\\
Fe     & \ion{}{11}	& 180.5 &   1.26\\
       & \ion{}{11} 	& 188.3 &   1.26\\
       & \ion{}{12} 	& 195.2 &   1.26\\
       & \ion{}{13} 	& 202.1 &   1.58\\
       & \ion{}{13} 	& 203.9 &   1.58\\
       & \ion{}{14} 	& 211.3 &   2.00\\
       & \ion{}{14} 	& 274.2 &   2.00\\
       & \ion{}{15} 	& 284.2 &   2.00\\
       & \ion{}{16} 	& 251.1 &   2.51\\
       & \ion{}{16} 	& 262.9 &   2.51\\
       & \ion{}{23} 	& 264.0 &   15.85\\
Ca     & \ion{}{17} 	& 192.8 &   5.01\\
\enddata
\label{tab:obsinfo}
\end{deluxetable}

\clearpage

\begin{figure*}[htbp!]
\centering
\includegraphics{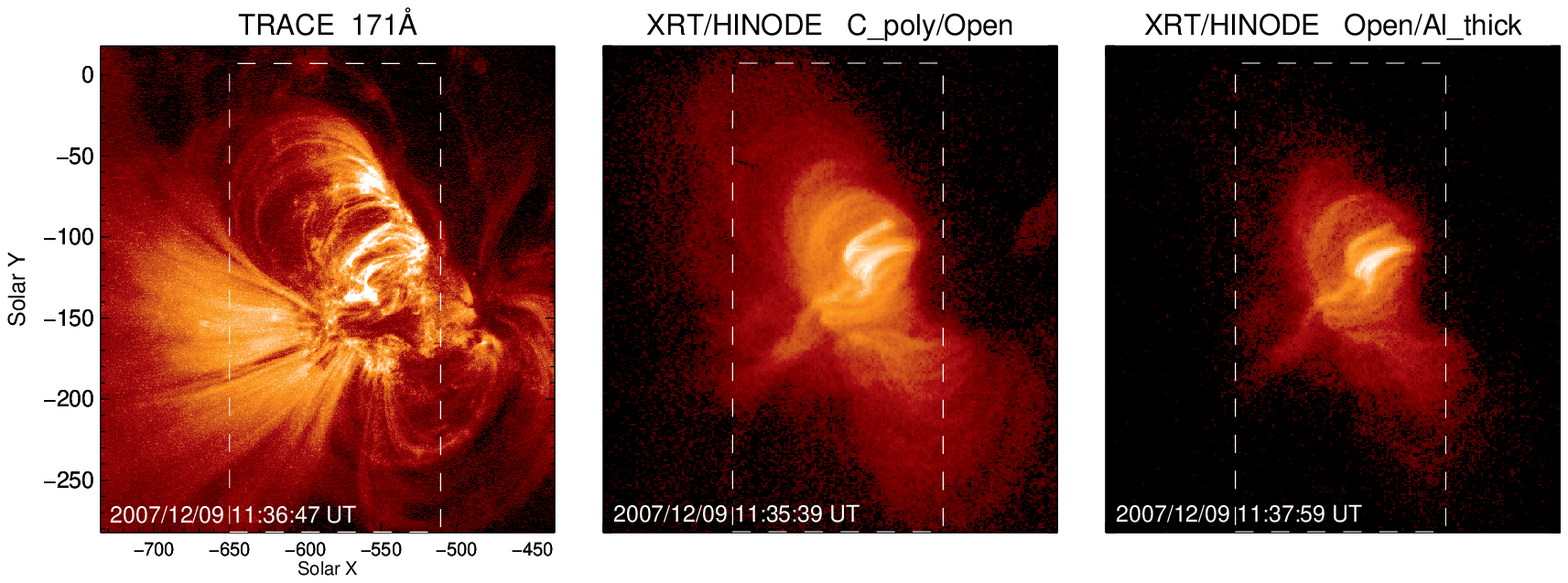}
\includegraphics{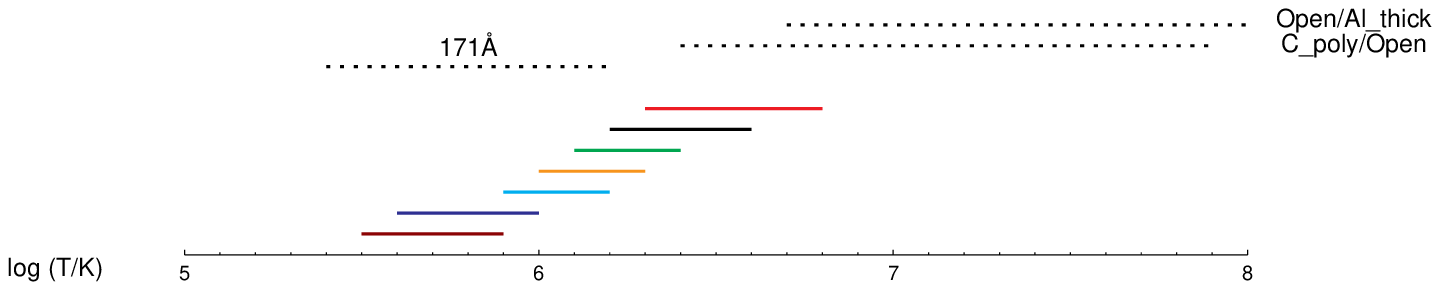}
\includegraphics{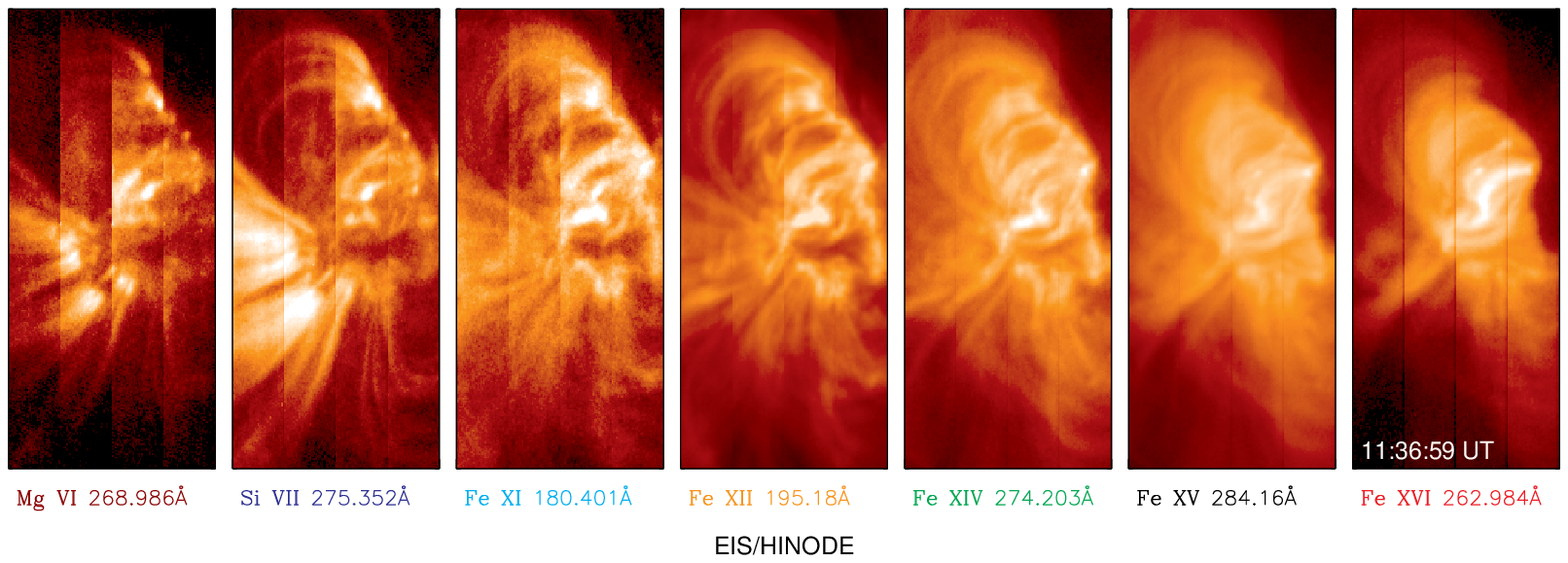}
\caption{Active region NOAA 10978 as seen by an EUV and an X-Ray imager ({\it TRACE} and XRT/{\it Hinode}), 
top panels, and as seen by EIS slot imaging. The center graph indicates the temperature response of the different
instruments.}
\label{fig:bigpanel}
\end{figure*}

\clearpage

\begin{figure}[htbp!]
\centering
\plotone{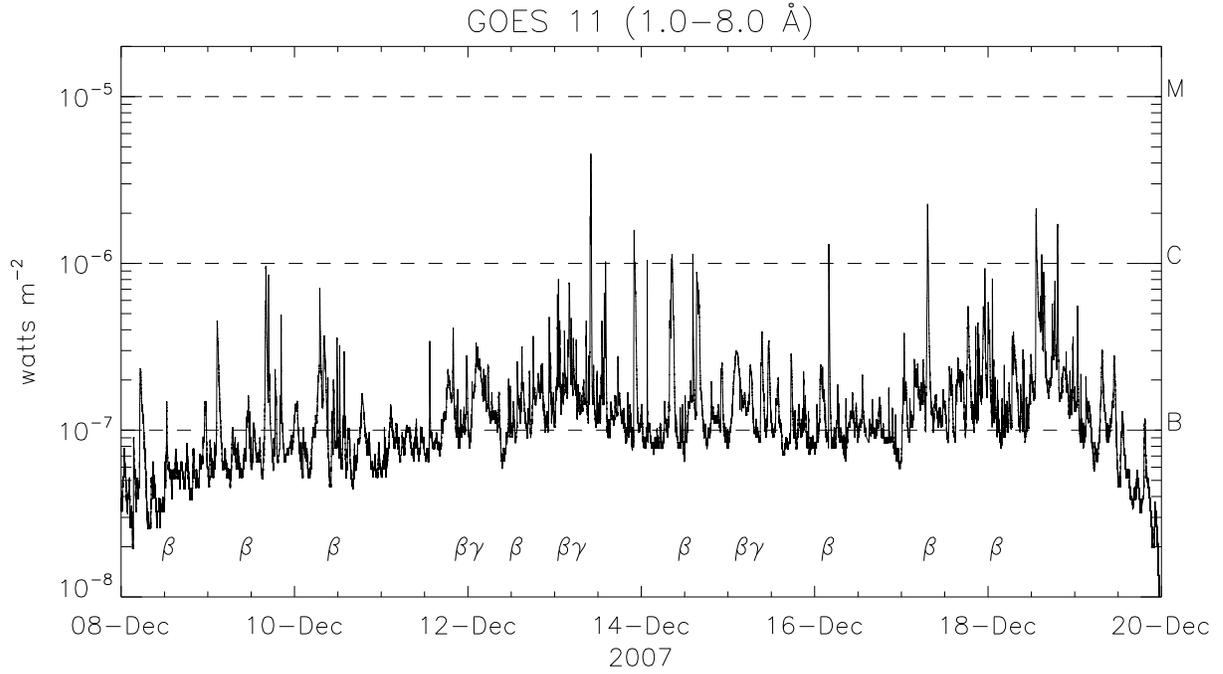}
\caption{GOES 1-8 \AA\ fluxes during the observing period. Flare class is indicated by the dashed lines
and M, C, B letters. Greek letters indicate AR10978 magnetic classification given by NOAA.}
\label{fig:goes}
\end{figure}

\clearpage
\begin{deluxetable}{clrlrlrlrl}
\footnotesize
\tablecolumns{10}
\tablecaption{Sample of loop lifetimes}
\tablewidth{0pt}
\rotate
\tablehead{
 	&  	&  \multicolumn{2}{c}{\ion{Fe}{15}}   & \multicolumn{2}{c}{\ion{Fe}{12}}	&  \multicolumn{2}{c}{\ion{Si}{7}}	   &  \multicolumn{2}{c}{\ion{Mg}{6}} 		   \\
No.	& Date	& Start Time &  Lifetime	& Start Time &  Lifetime	& Start Time &  Lifetime	& Start Time &  Lifetime}
\startdata
1   & 2007/12/09   & 13:53:35 & 770s & 14:06:24  & 630s 	& 14:16:55	&  770s		& 14:18:05 & 770s \\
2   & 2007/12/09   & 14:36:45 & 770s & 14:43:46  & 560s 	& 14:44:55	&  560s		& 14:44:55 & 560s \\
3   & 2007/12/10   & 13:54:53 & 1820s & 14:10:04  & 1050s 	& 14:17:04	&  840s		& 14:19:23 & 770s \\
4   & 2007/12/10   & 13:56:02 & 1261s & 14:05:23  & 980s 	& 14:06:34	&  840s		& 14:07:43 & 770s \\
5   & 2007/12/11   & 07:23:45 & 1470s & 07:38:55  & 840s 	& 07:43:35	&  1121s	& 07:44:45 & 981s \\
6   & 2007/12/11   & 20:49:40 & 3431s & 21:24:40  & 1540s 	& 21:49:10	&  1121s	& 21:52:41 & 1190s \\
7   & 2007/12/12   & 19:51:51 & 1961s & 20:18:41  & 1330s 	& 20:28:01	&  1332s	& 20:29:12 & 1191s \\
8   & 2007/12/12   & 20:42:02 & 1786s & 21:05:24  & 1226s 	& 21:22:20	&  1190s	& 21:24:39 & 1120s \\
9   & 2007/12/13   & 21:04:13 & 4165s & 21:39:16  & 1993s 	& 22:27:40	&  1820s	& - & - \\
\enddata
\label{tab:looplifetimes}
\end{deluxetable}

\clearpage

\thispagestyle{empty}
\setlength{\voffset}{-15mm}
\begin{figure*}[htbp!]
\centering
\includegraphics[angle=90,width=17cm]{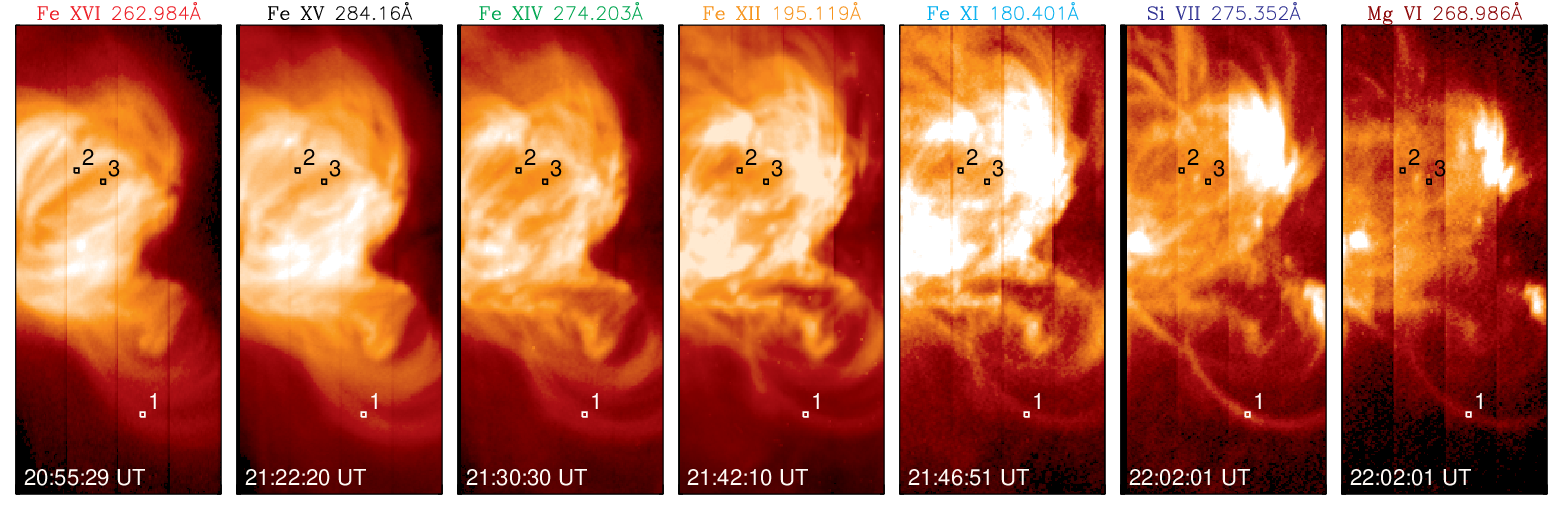}
\includegraphics[angle=90,width=5.4cm]{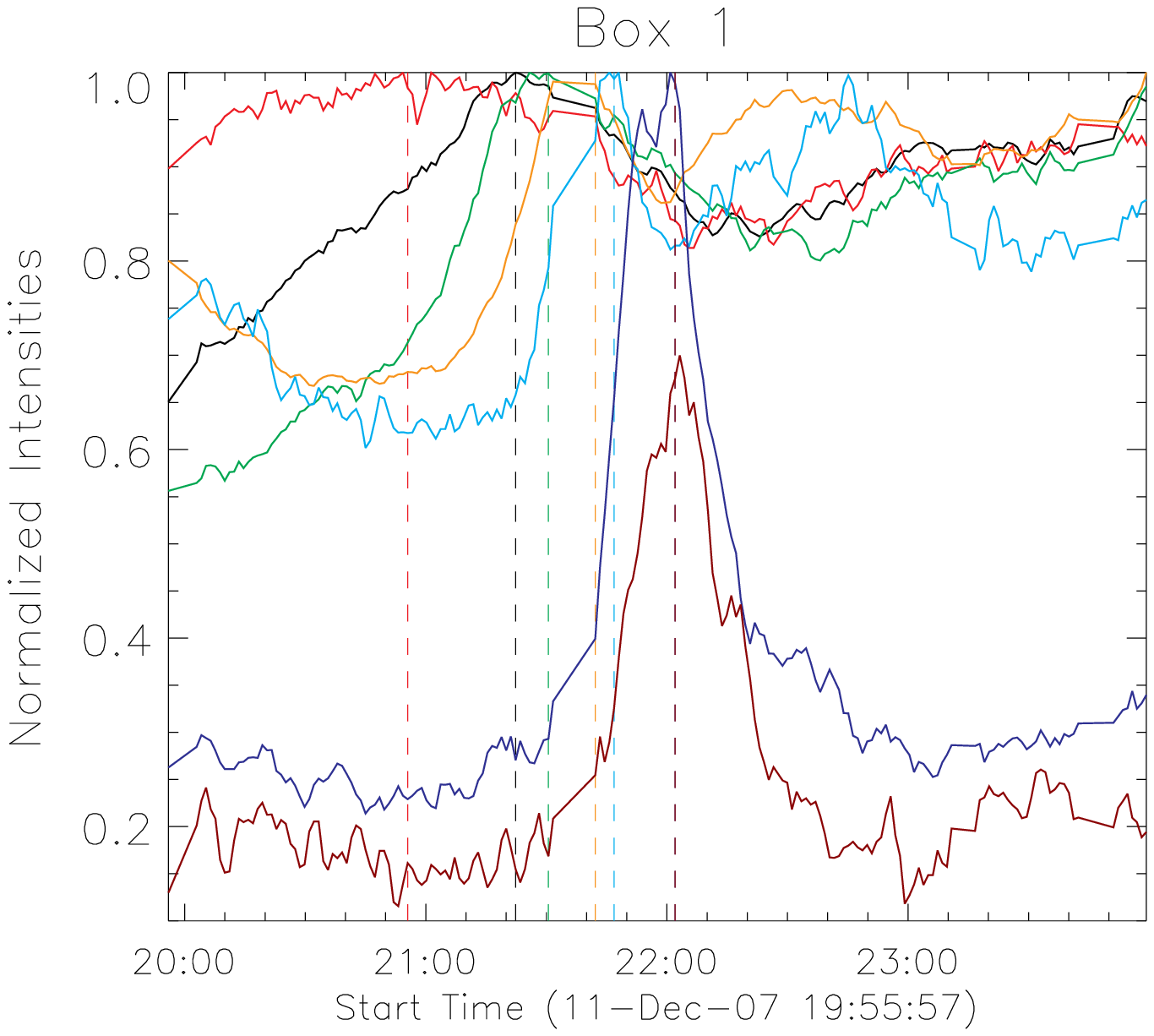}
\includegraphics[angle=90,width=5.4cm]{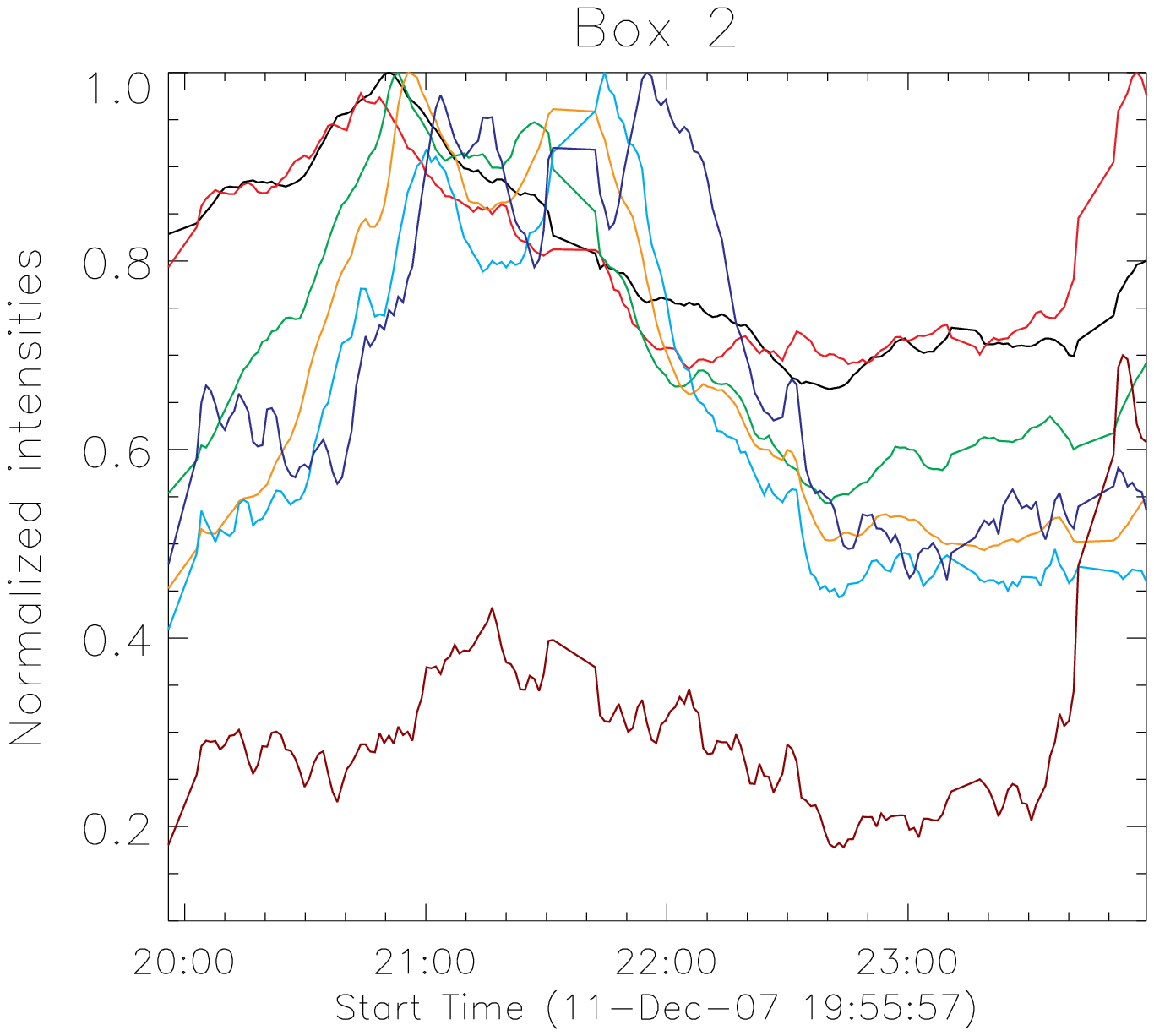}
\includegraphics[angle=90,width=5.4cm]{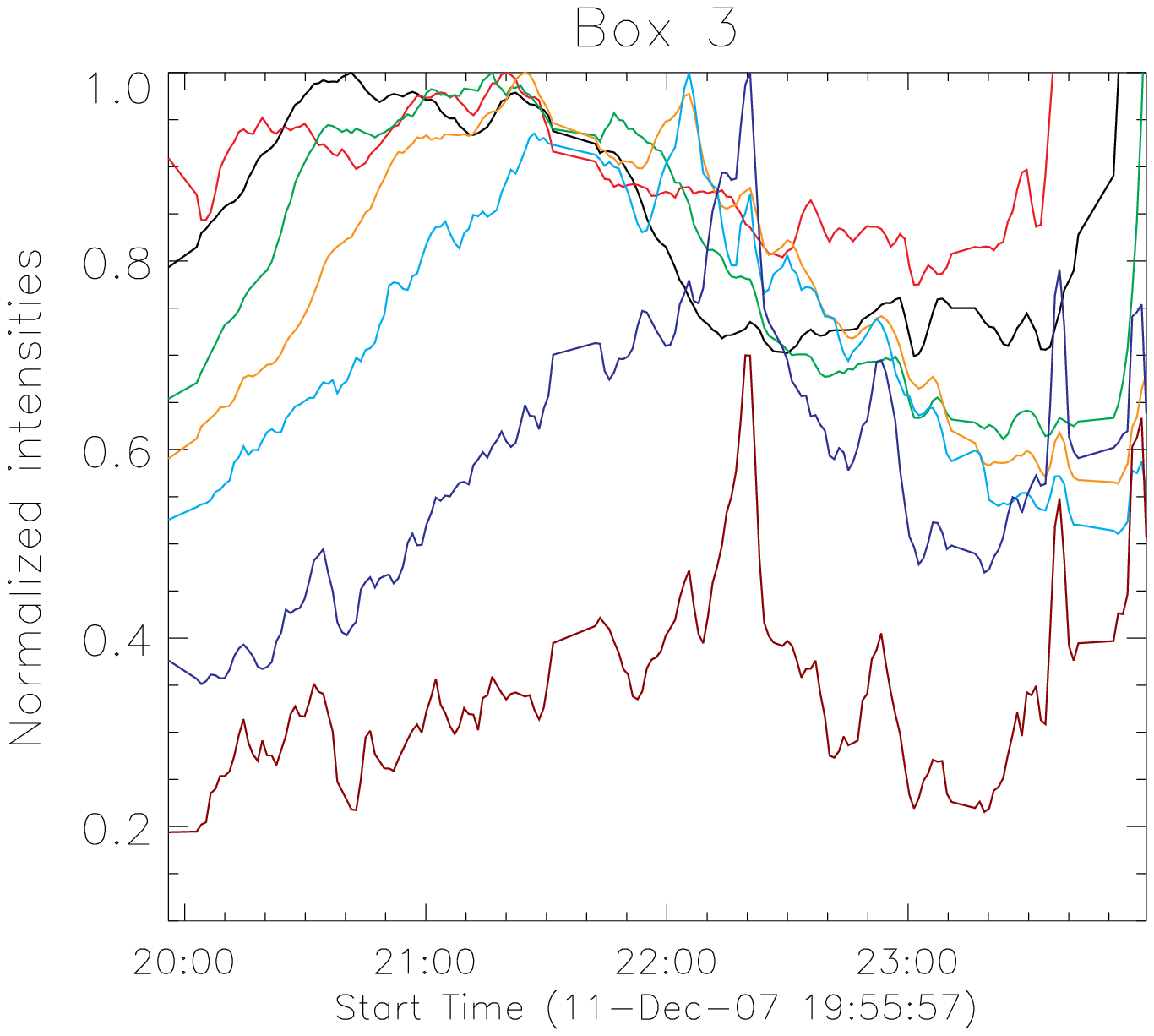}
\includegraphics[angle=90,width=17cm]{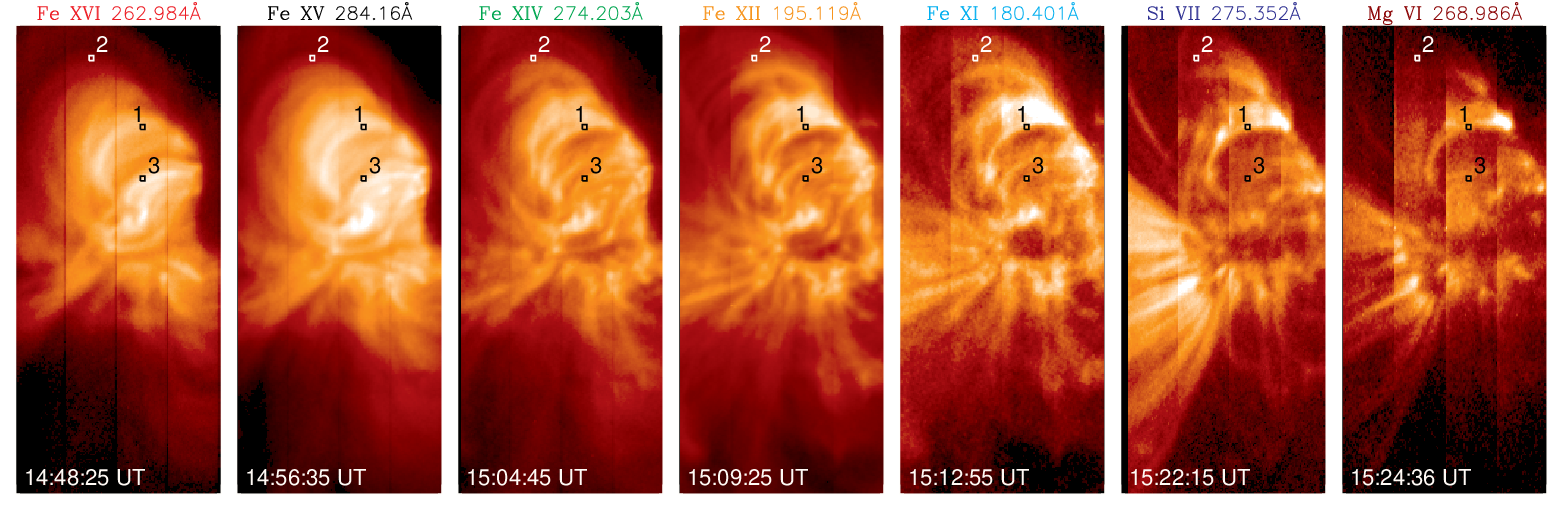}
\includegraphics[angle=90,width=5.4cm]{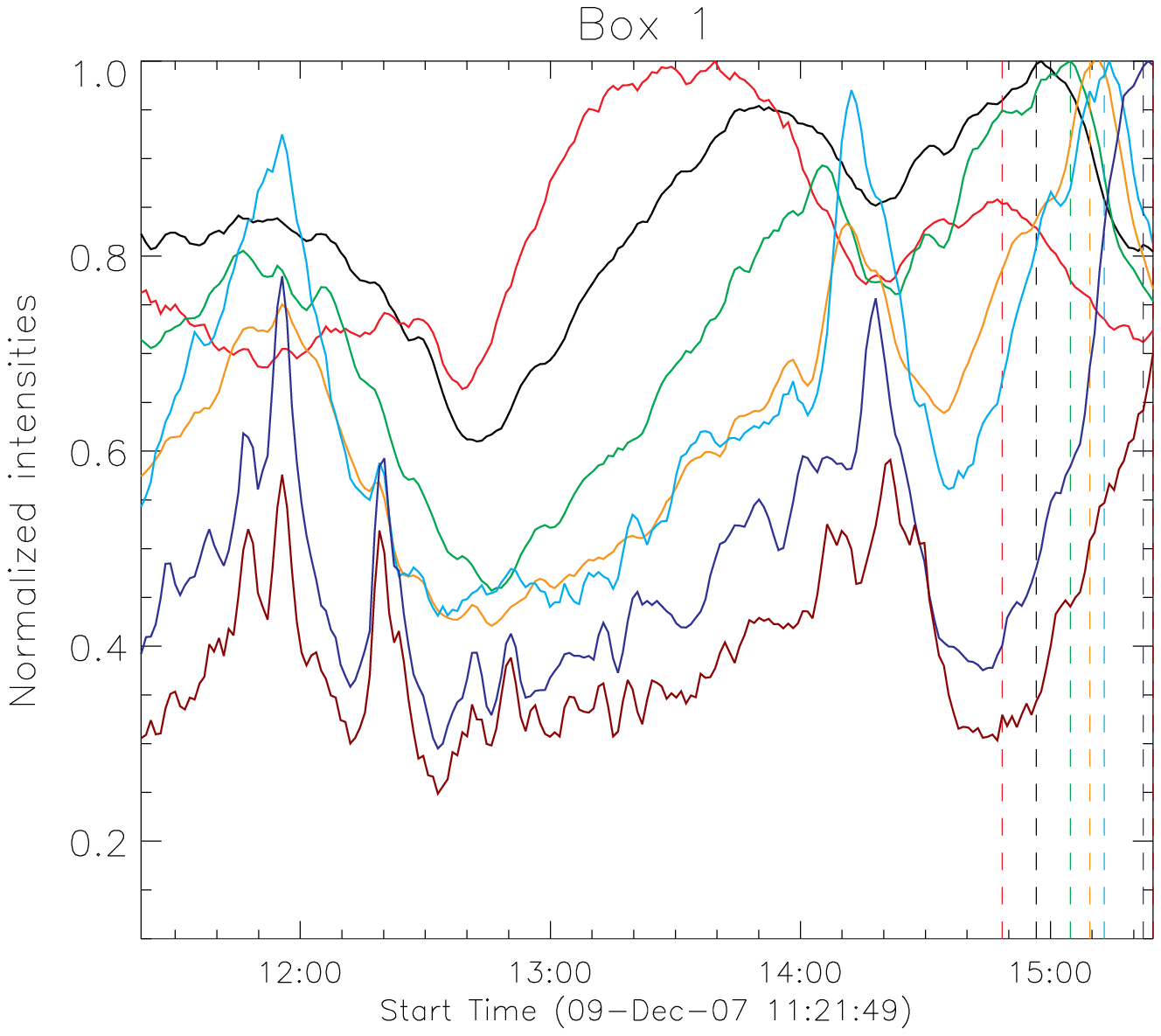}
\includegraphics[angle=90,width=5.4cm]{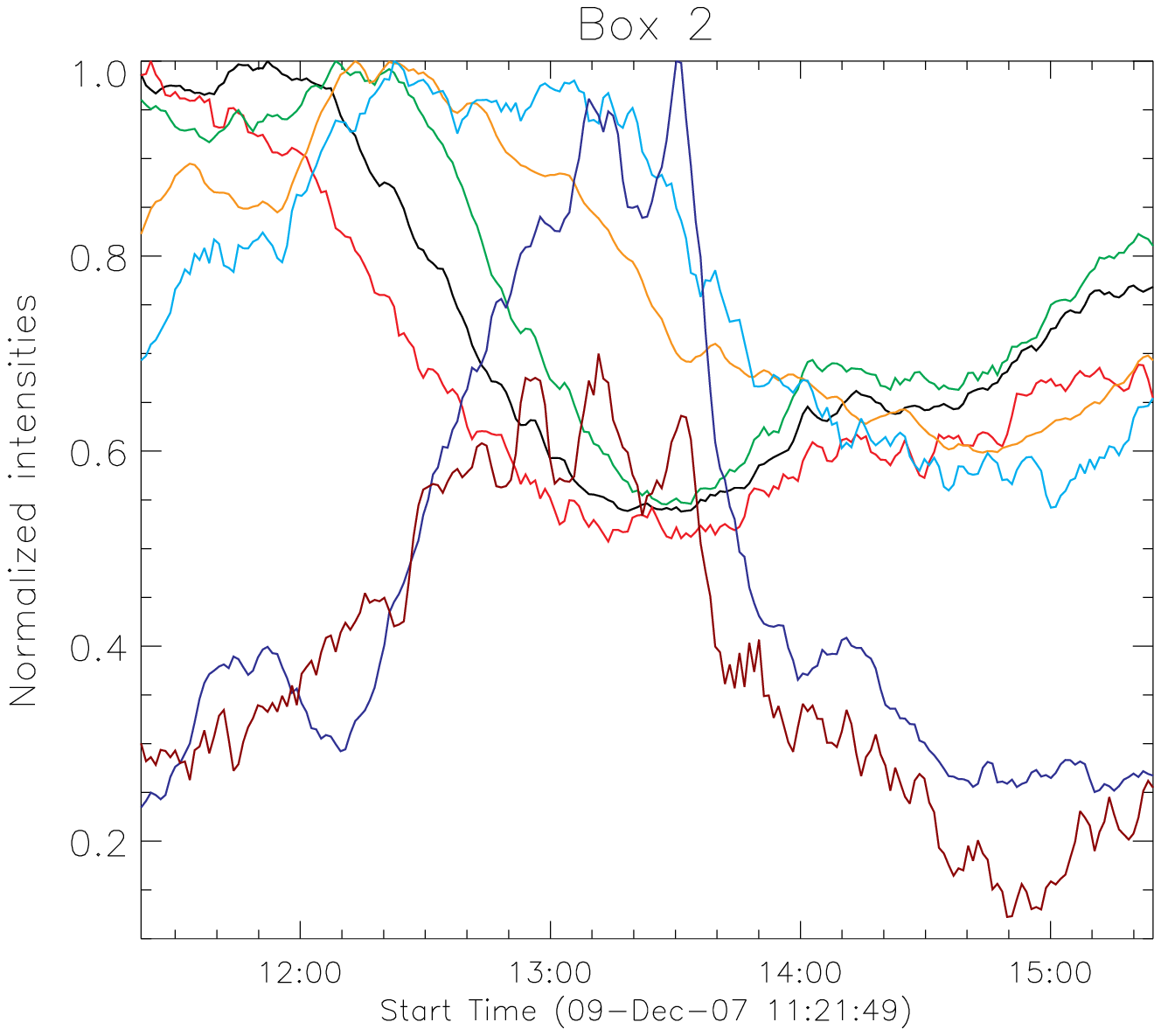}
\includegraphics[angle=90,width=5.4cm]{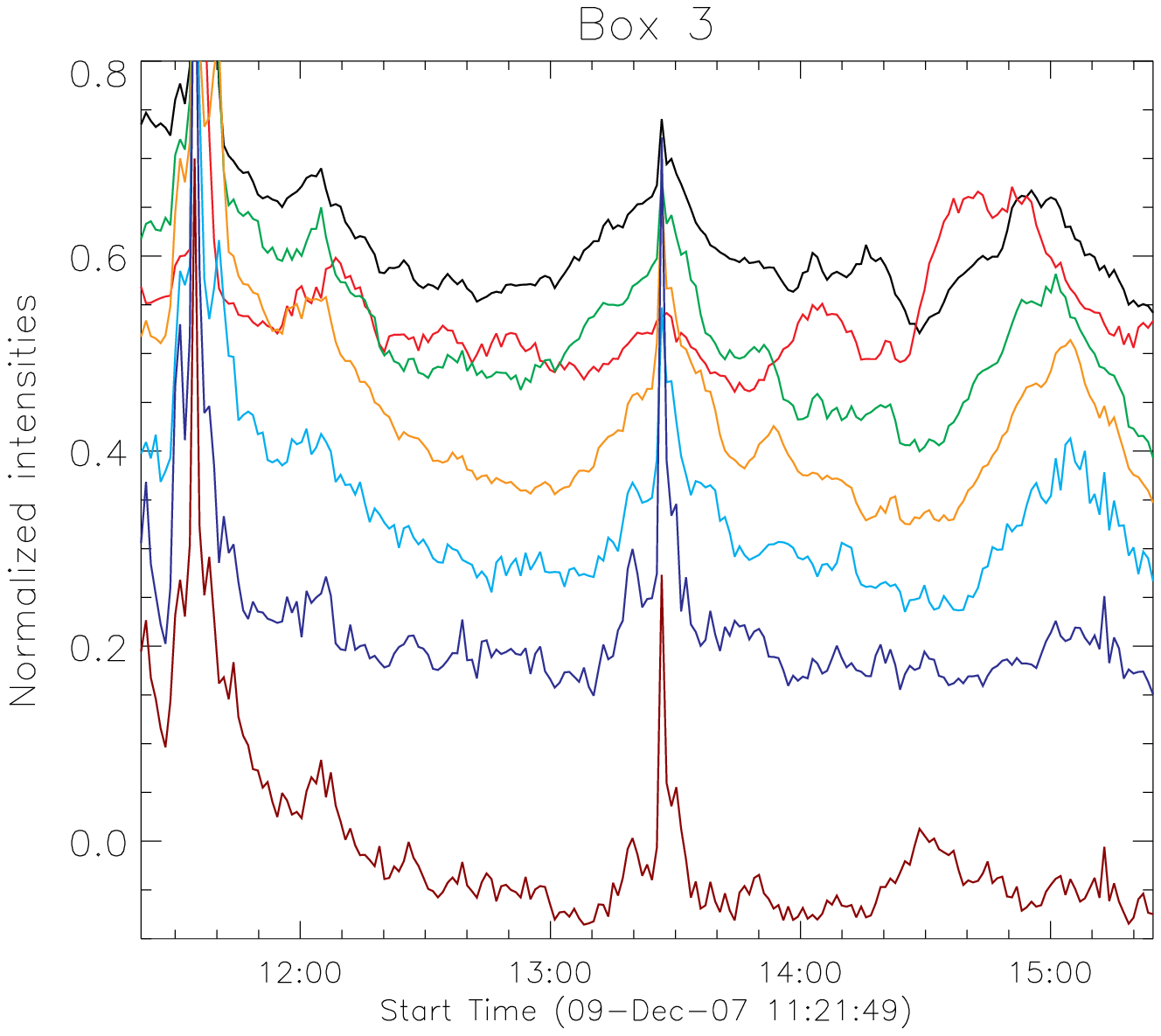}
\caption{Compact multi-temperature loops. Normalized intensity lightcurves for several spectral lines at three different
locations: boxes 1, 2 and 3. Top half corresponds to December 11 2007 observing time and bottom half to December 9 2007. 
Times of slot images correspond to dashed lines on box 1 lightcurves, i.e. peak in intensity. Integration boxes are 
3$\times$3 pixels. For the sake of clarity, all the curves (but box 3 in December 9)  have been smoothed with a boxcar
average of 3 data points in time and an absolute value of 0.3 has been subtracted of the \ion{Mg}{6} curves.}
\label{fig:loops_lightc}
\end{figure*}

\clearpage

\begin{figure*}[htbp!]
\centering
\includegraphics[width=7.5cm]{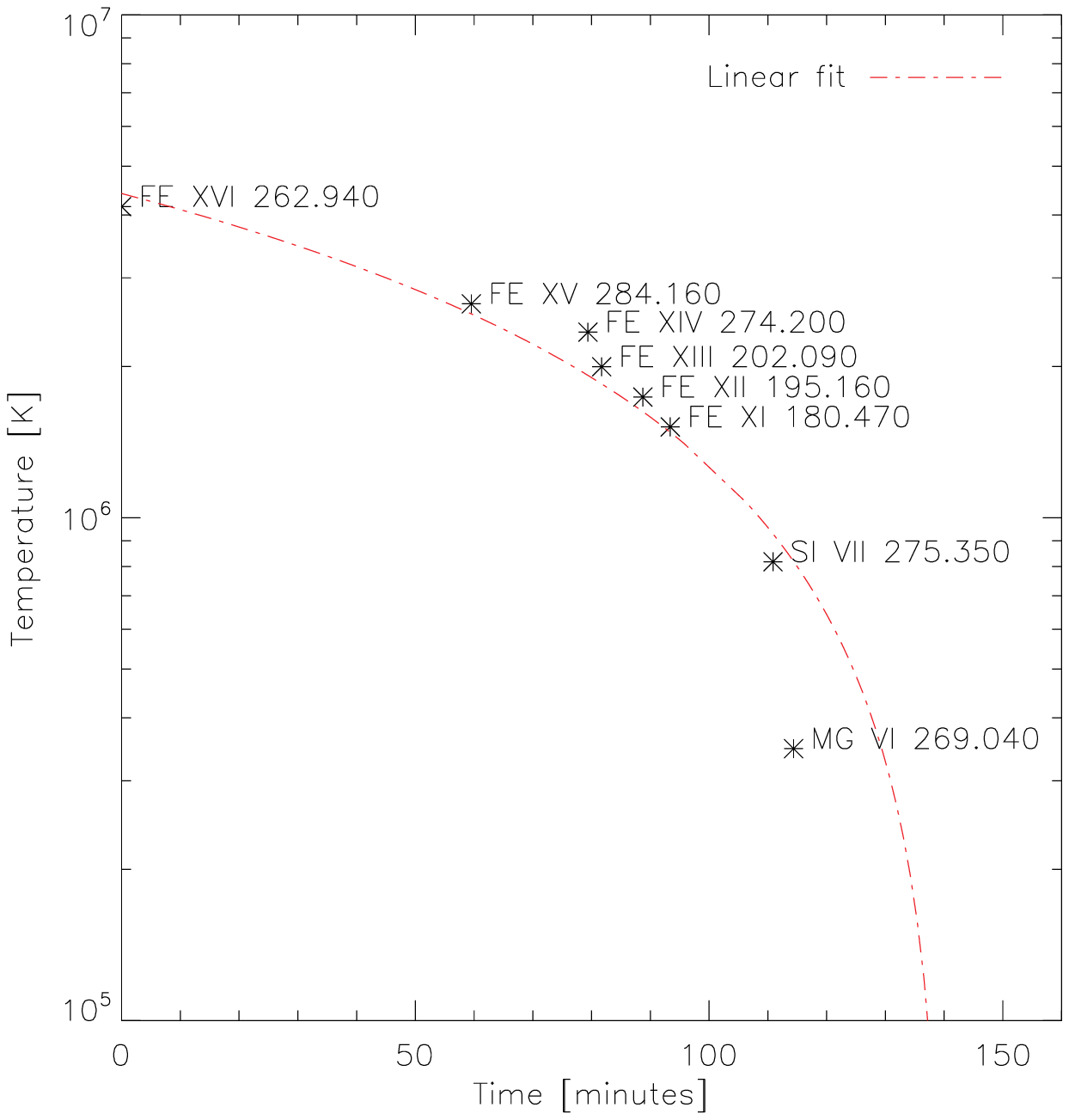}
\includegraphics[width=8.5cm]{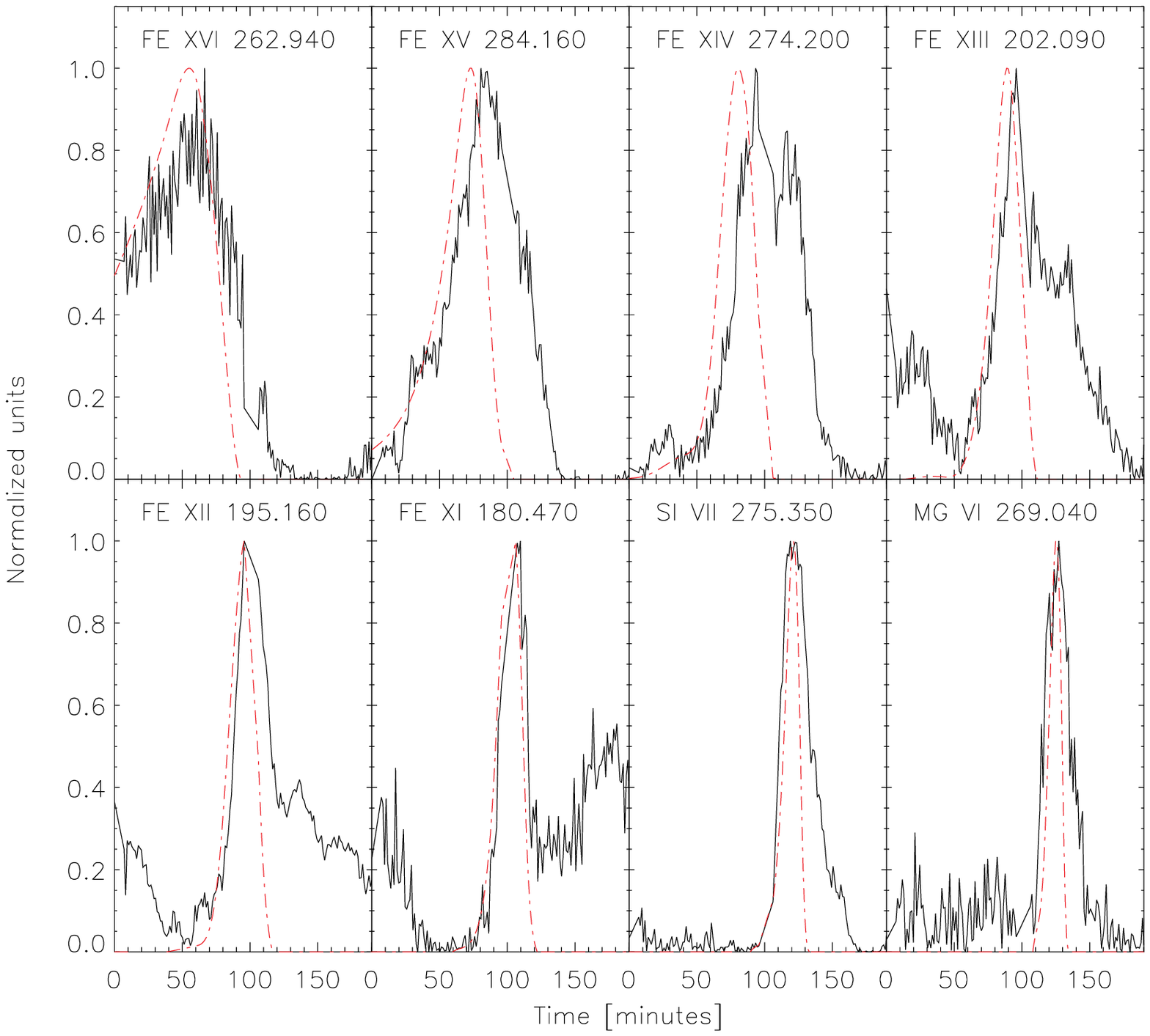}
\caption{Left panel: Temperature decay as a function of time for the loop under box 1 on December 11. Right panel: normalized 
background subtracted intensity lightcurves in solid black and variation of the contribution function as function of time 
and temperature change in red dot-dashed line.} 
\label{fig:tdecay}
\end{figure*}

\clearpage
\setlength{\voffset}{0mm}

\begin{figure*}[htbp!]
\centering
\includegraphics[angle=90,width=17cm]{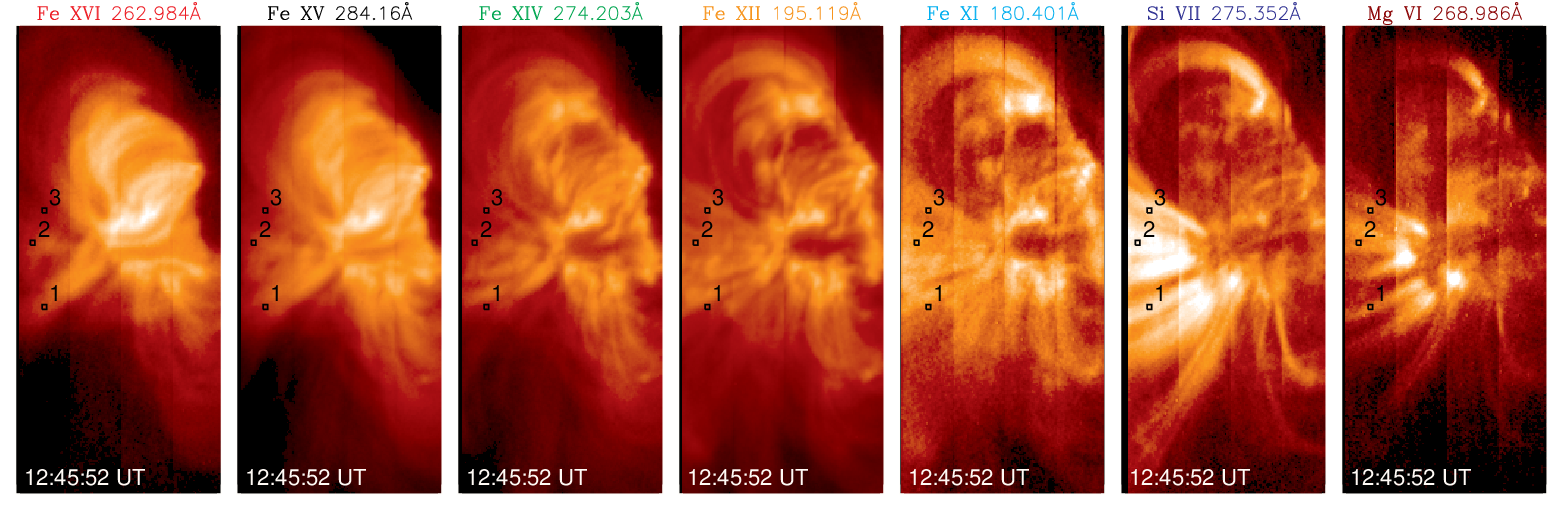}
\includegraphics[angle=90,width=5.4cm]{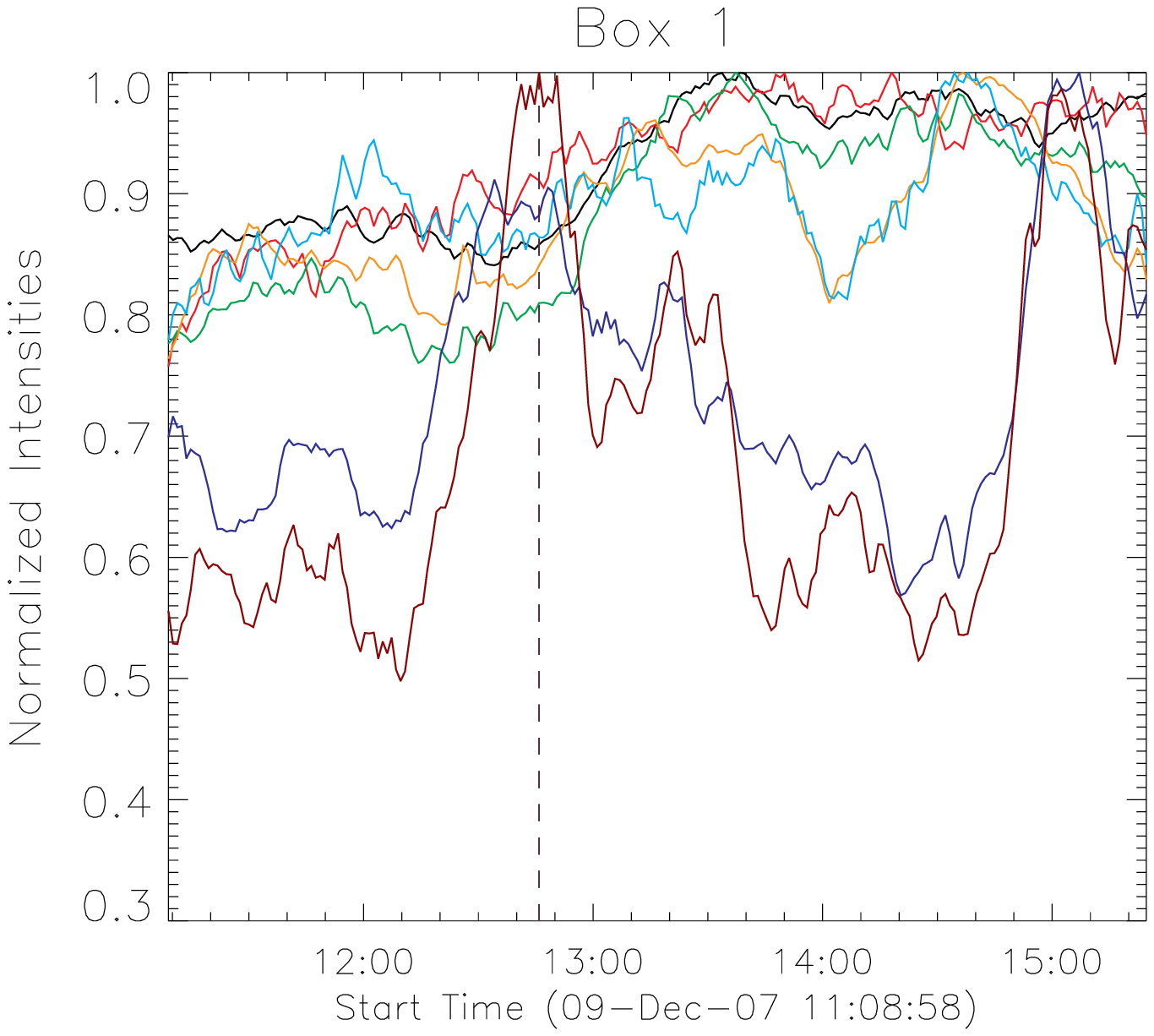}
\includegraphics[angle=90,width=5.4cm]{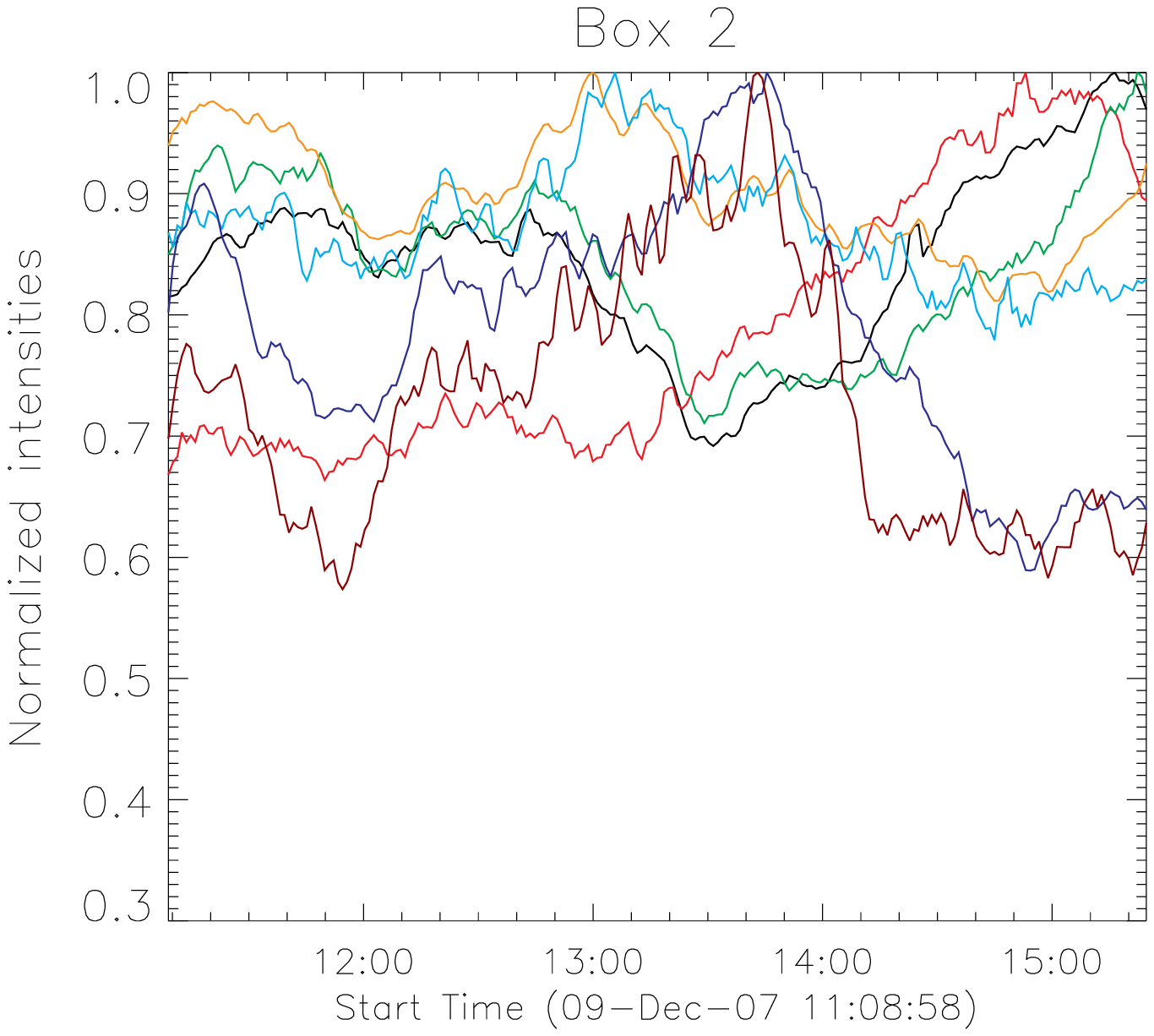}
\includegraphics[angle=90,width=5.4cm]{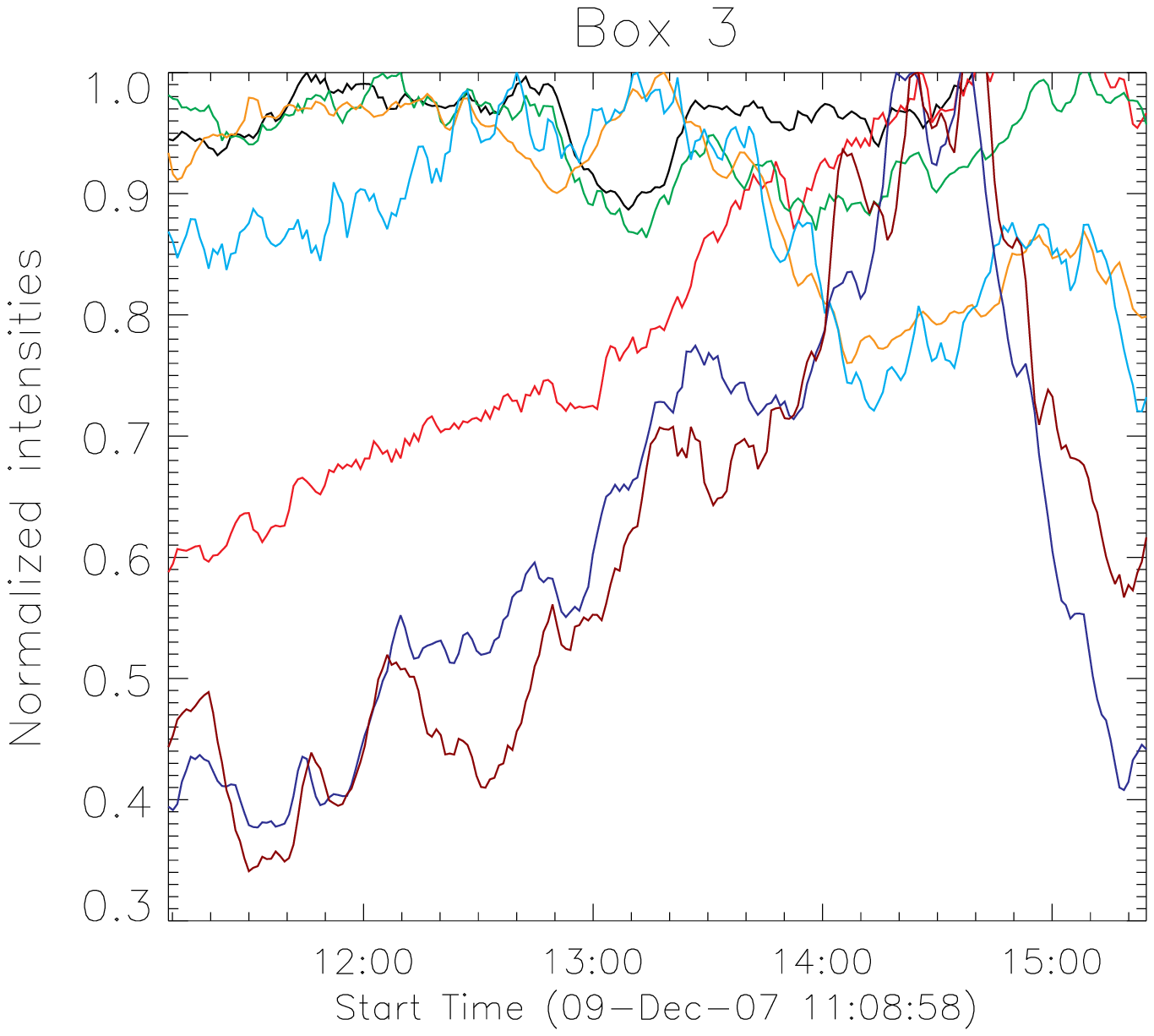}
\includegraphics[angle=90,width=17cm]{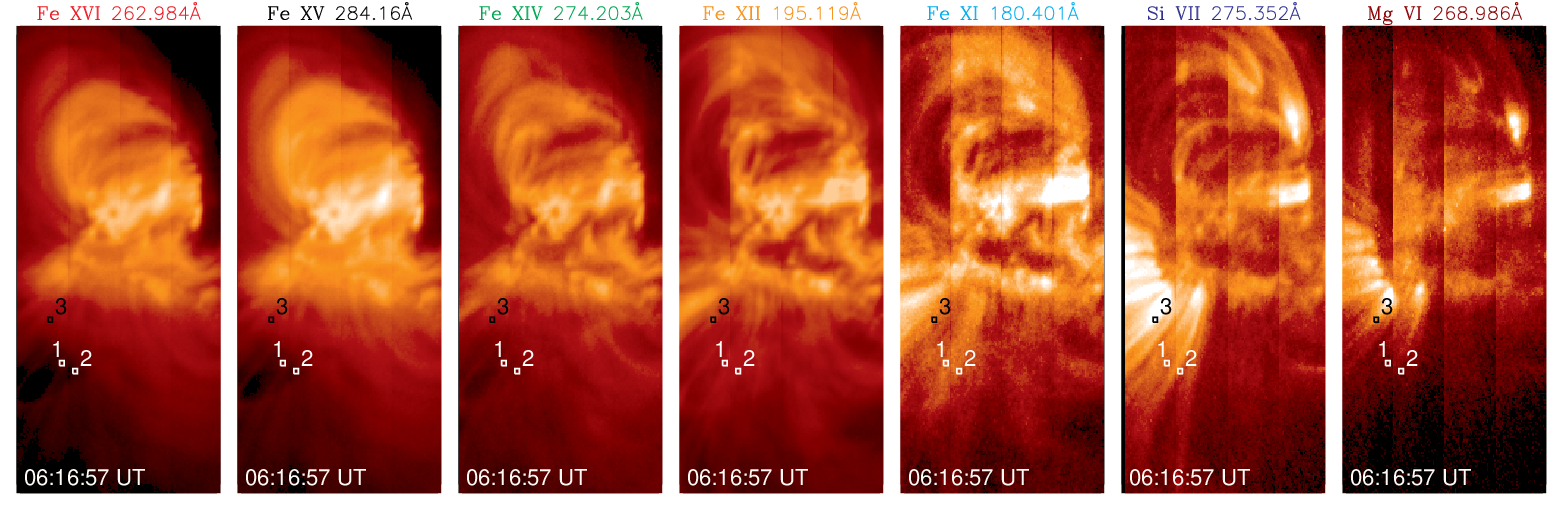}
\includegraphics[angle=90,width=5.4cm]{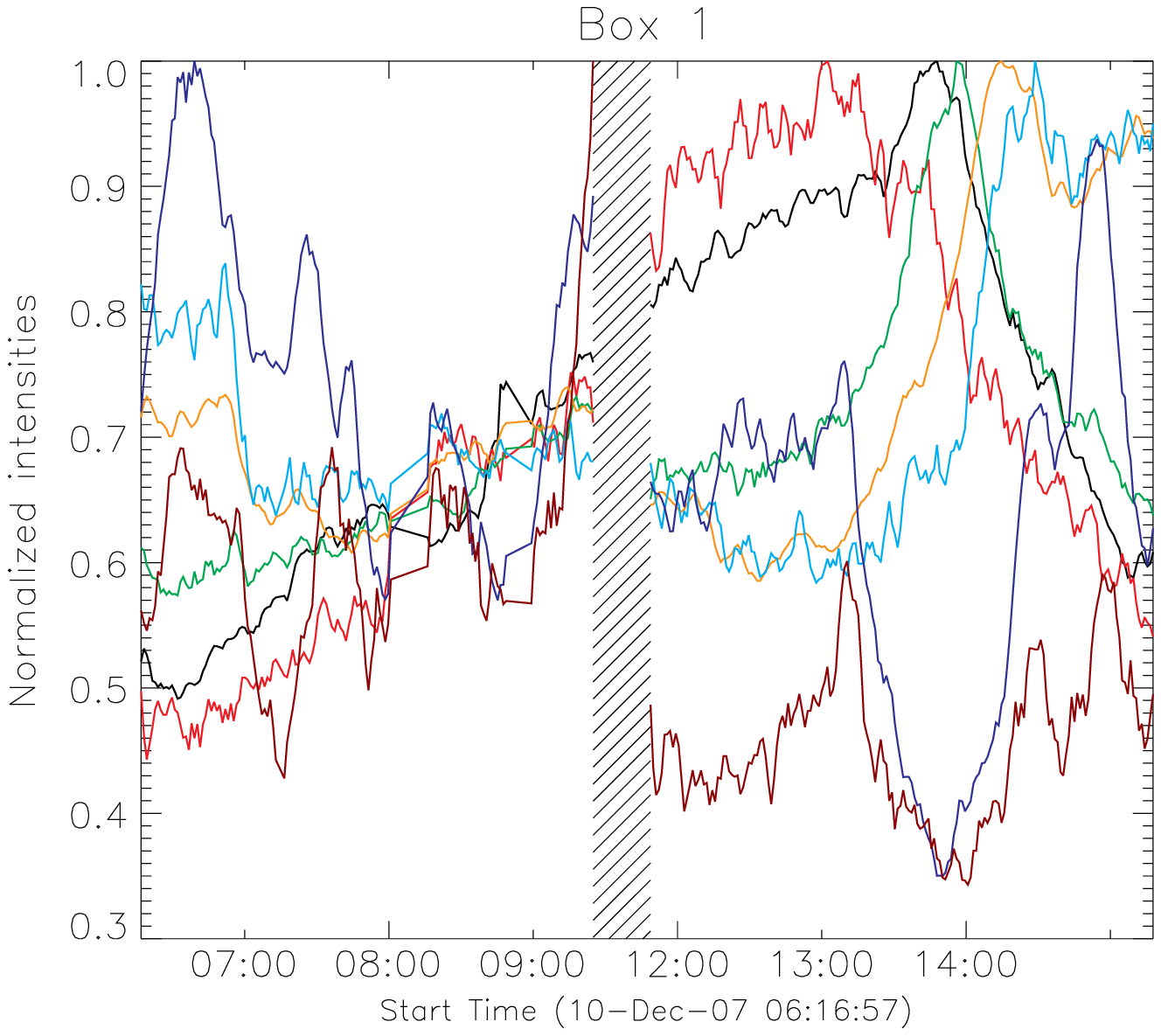}
\includegraphics[angle=90,width=5.4cm]{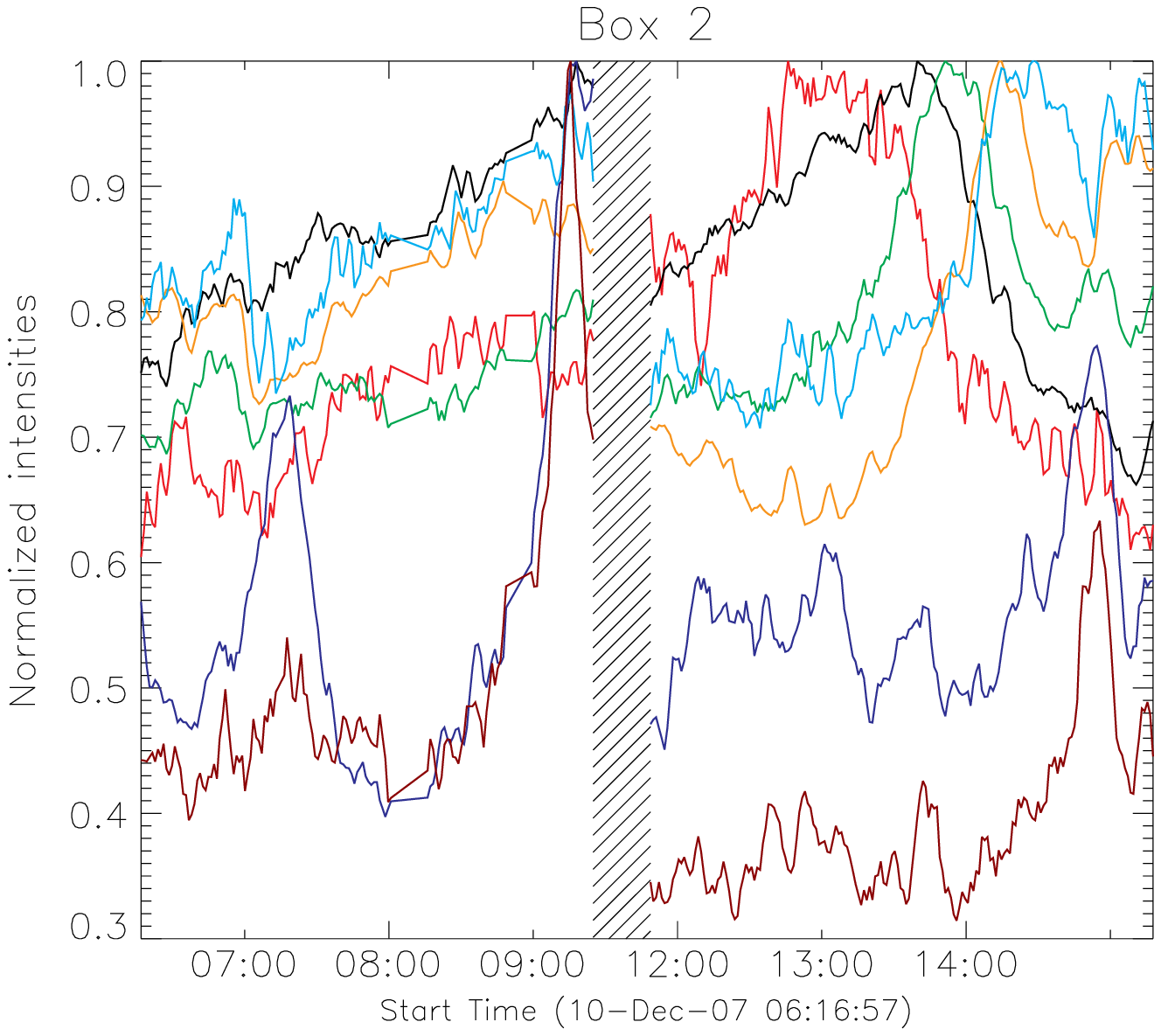}
\includegraphics[angle=90,width=5.4cm]{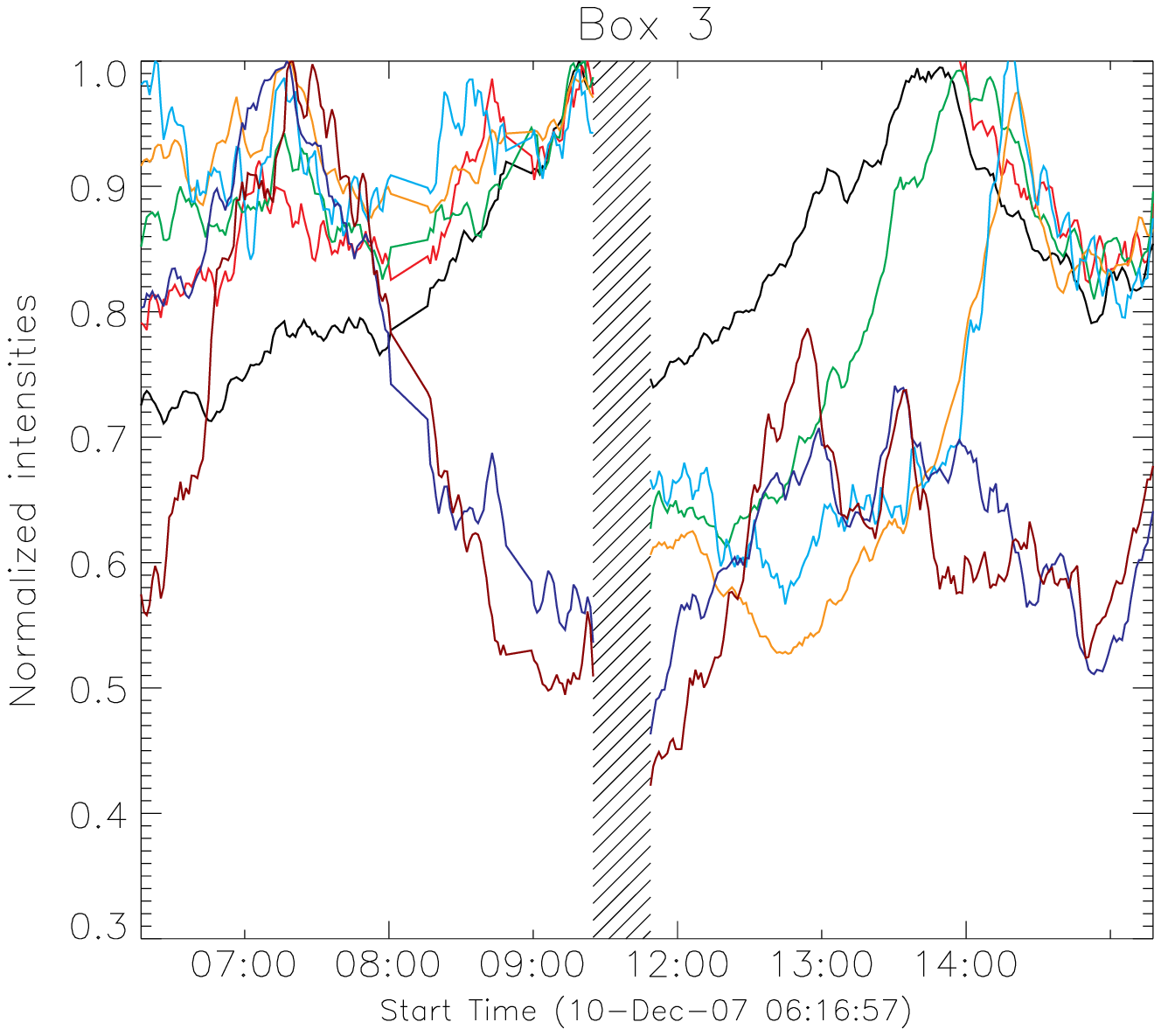}
\caption{Peripheral cool loops. Filled pattern indicates there is a data gap of approximately 2.4 hours. Notice times on
the abscissa axis. For the sake of clarity, all the curves have been smoothed with a boxcar average of 3 data points in time.
There is very little correspondence between intensity changes at transition region temperatures and the corona.}
\label{fig:coolloops_lightc}
\end{figure*}

\clearpage

\begin{figure*}[htbp!]
\centering
\includegraphics[angle=90,width=17cm]{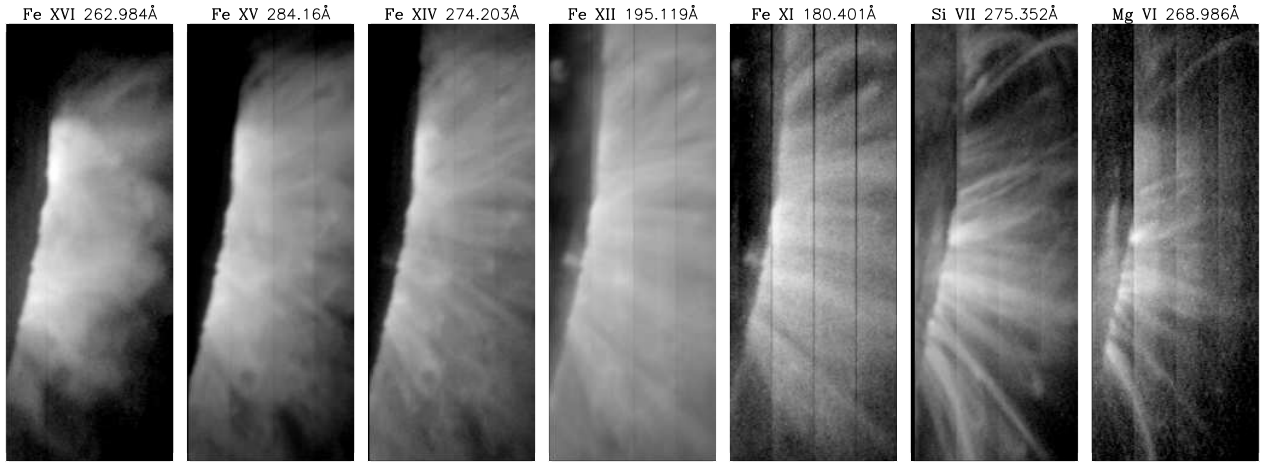}
\caption{Active region loops as seen on projection over the West limb in various 
different spectral lines, covering a temperature range of 2.5 MK - 0.4 MK. Time: 07:09 UT, December 19 2007.}
\label{fig:loops_offlimb}
\end{figure*}

\clearpage

\begin{figure*}[htbp!]
\centering
\includegraphics[width=5.4cm]{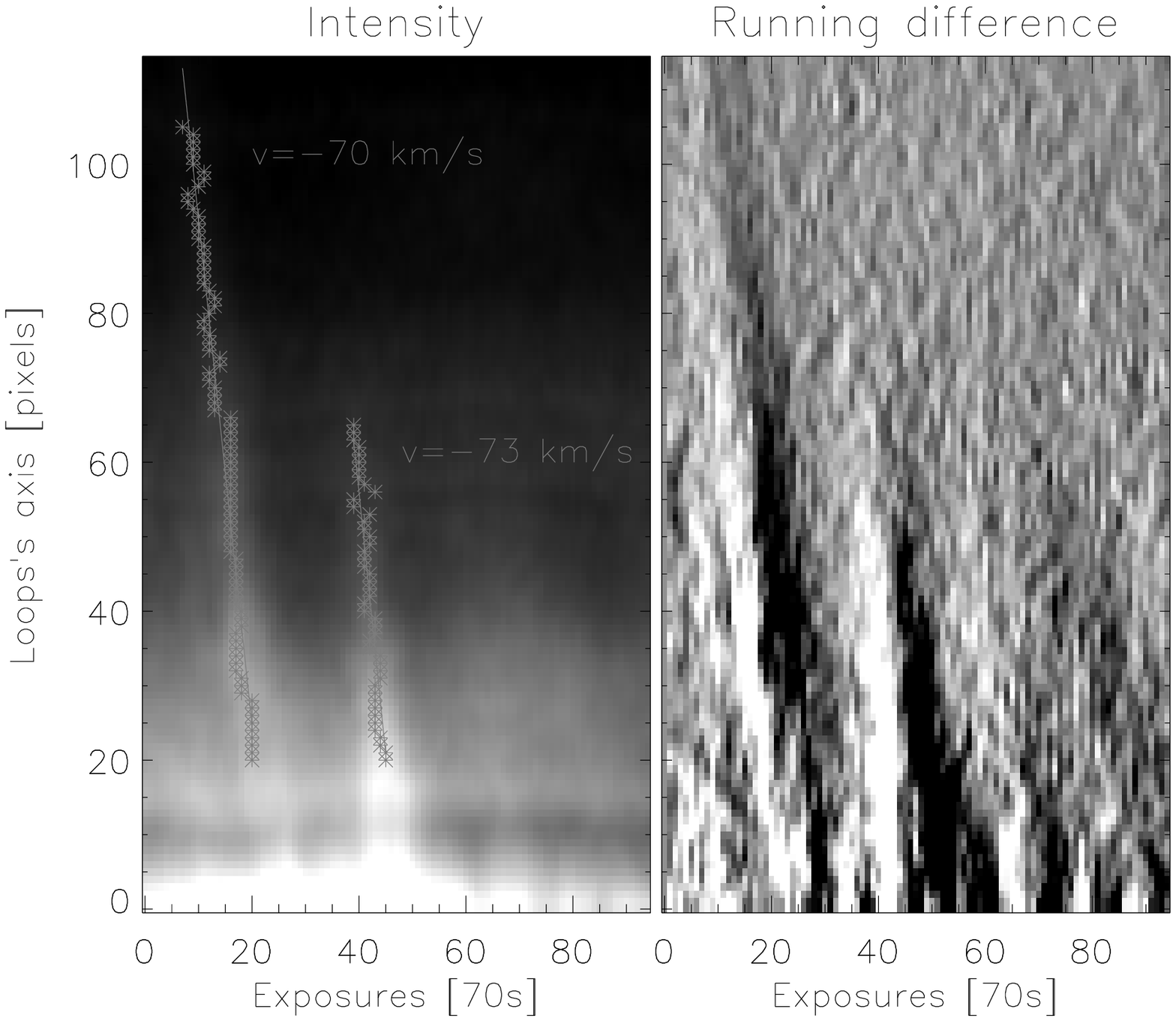}
\includegraphics[width=5.4cm]{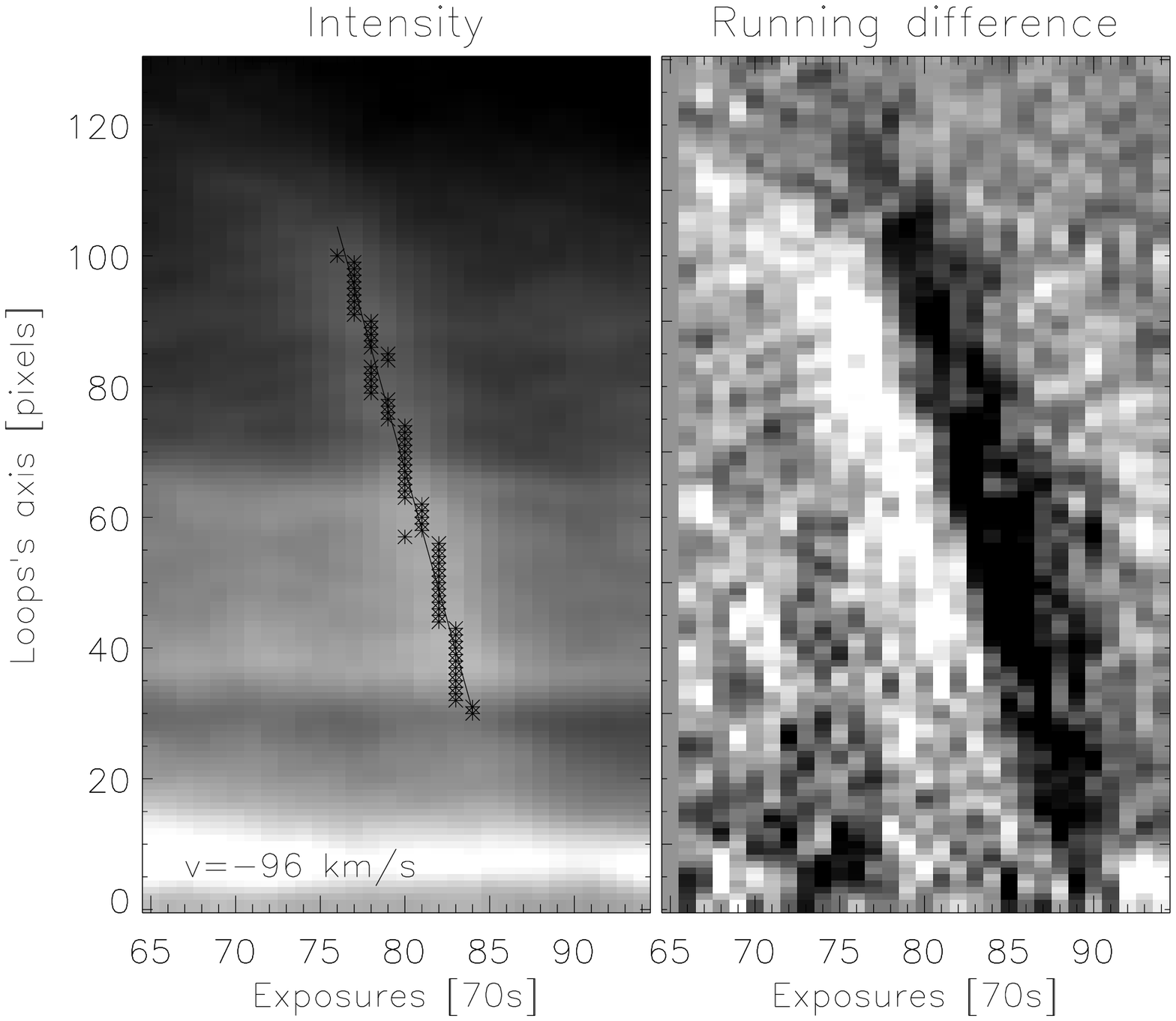}
\includegraphics[width=5.4cm]{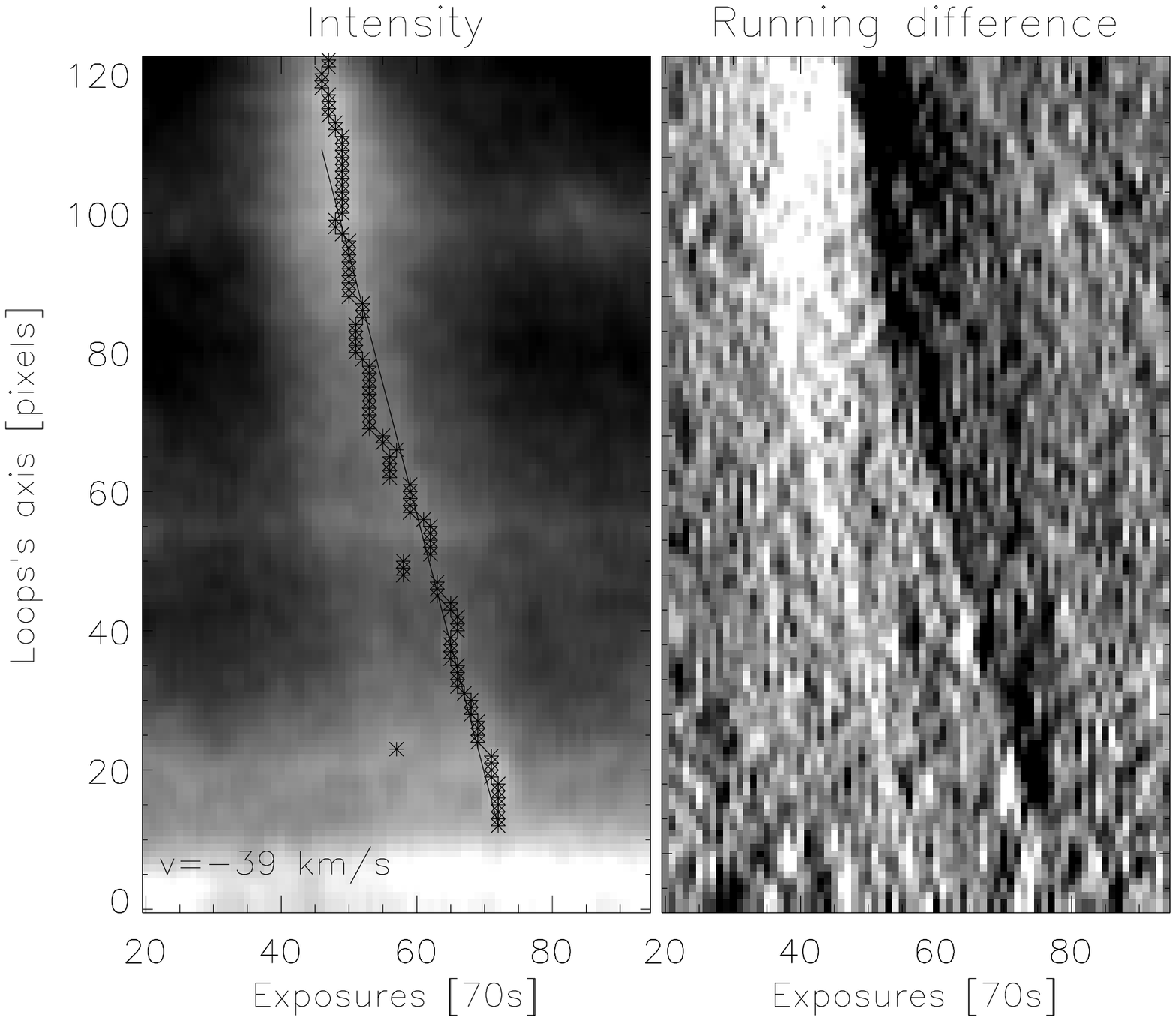}
\vspace{0.2cm}
\caption{Downflows along the legs of three different \ion{Mg}{6} off-limb loops. Asterisks indicate the maximum of the 
intensity in the lightcurve at a given location along the loop's axis. Solid lines are linear fits to the points. The 
slope of the curve is the associated velocity. Exposure 0 corresponds to time 05:55 UT, December 19 2007.} 
\label{fig:loops_flows}
\end{figure*}

\clearpage

\begin{figure*}[htbp!]
\centering
\includegraphics{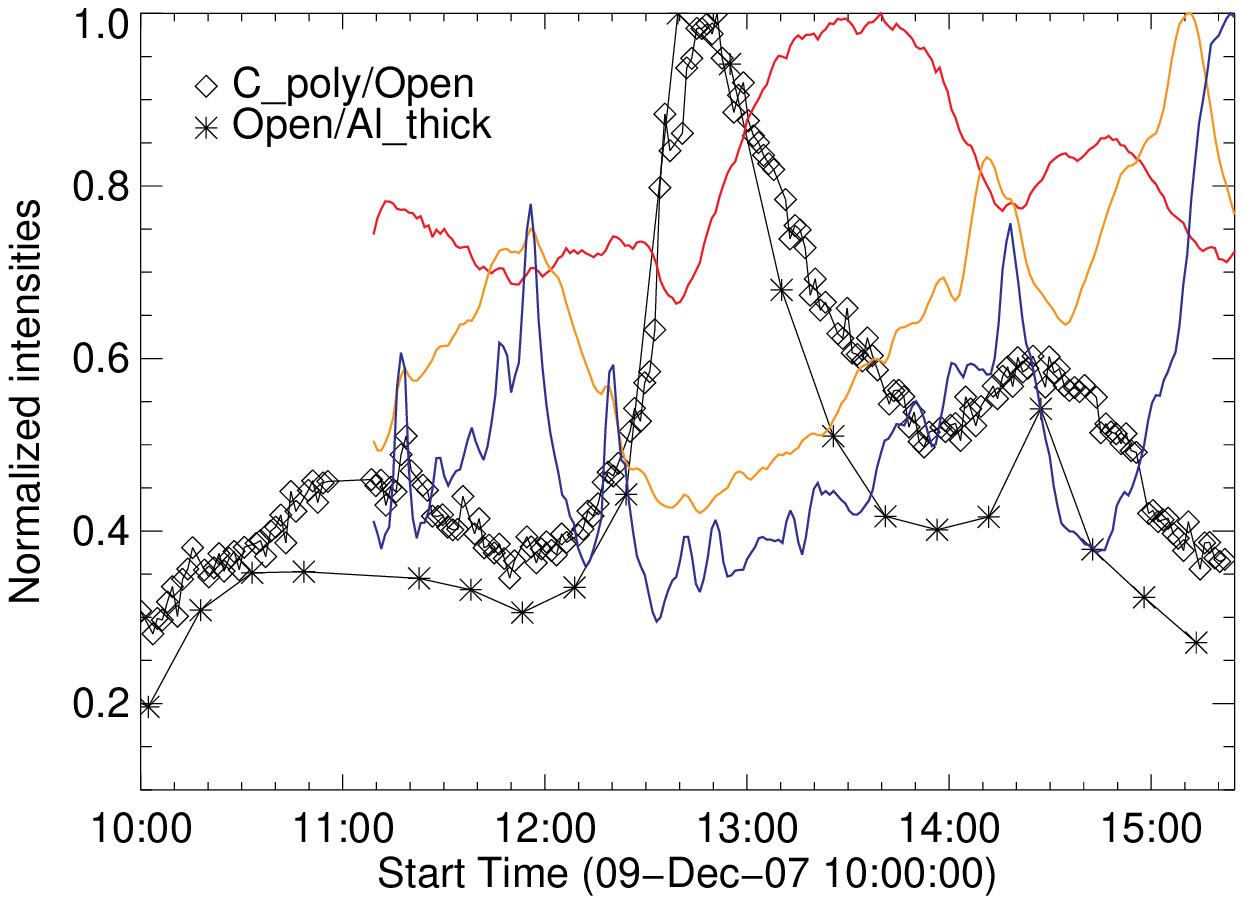}
\includegraphics{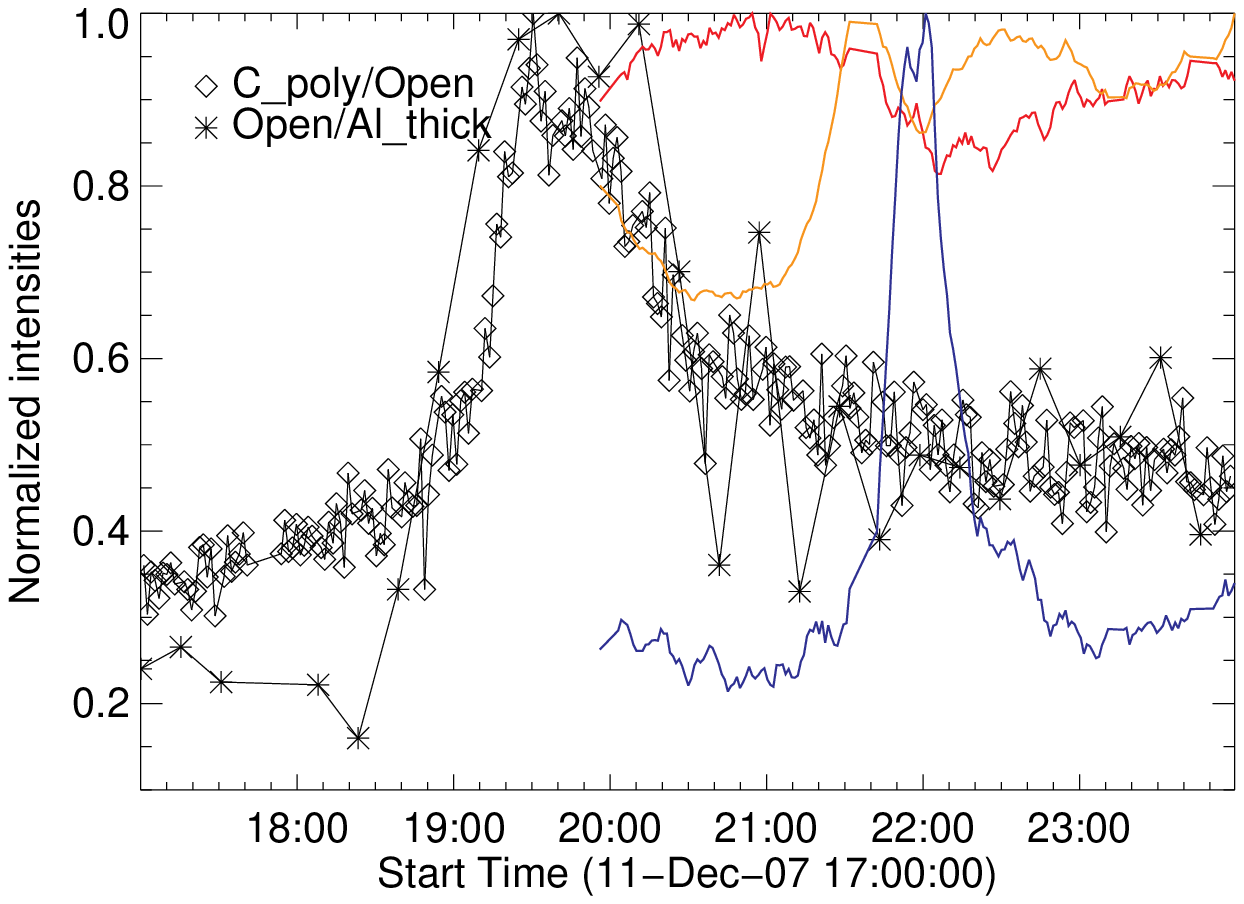}
\caption{XRT Box 1 lightcurves for two filter combinations (diamonds and asterisks) on December 9 and 11. 
\ion{Fe}{16}, \ion{Fe}{12} and \ion{Si}{7} lightcurves also shown for comparison with  Fig~\ref{fig:loops_lightc}. Same color 
coding as that figure.} 
\label{fig:xrt_eis}
\end{figure*}

\end{document}